\documentclass[review]{elsarticle}
\usepackage{graphicx}
\usepackage{subfig}
\usepackage{ulem}
\usepackage{amsmath}
\usepackage{xcolor}
\usepackage{hyperref}
\usepackage[capitalise]{cleveref}
\usepackage[top=1.1in, bottom=1.1in, left=1.0in, right=1.0in]{geometry}
\usepackage{lineno,hyperref}
\modulolinenumbers[5]
\usepackage{bm}
\usepackage{graphicx}
\usepackage{graphics}
\usepackage{amsmath}
\usepackage{multirow}
\usepackage{booktabs}
\usepackage{cases}

\begin{document}
\title{Discrete unified gas kinetic scheme for multiscale heat transfer with arbitrary temperature difference}
\author[add1]{Chuang Zhang}
\ead{zhangcmzt@hust.edu.cn}
\author[add1]{Zhaoli Guo\corref{cor1}}
\ead{zlguo@hust.edu.cn}

\cortext[cor1]{Corresponding author}
  \address[add1]{State Key Laboratory of Coal Combustion, Huazhong University of Science and Technology,Wuhan, 430074, China}
\begin{abstract}

In this paper, a finite-volume discrete unified gas kinetic scheme (DUGKS) based on the non-gray phonon transport model is developed for multiscale heat transfer problem with arbitrary temperature difference.
Under large temperature difference, the phonon Boltzmann transport equation (BTE) is essentially multiscale, not only in the frequency space, but also in the spatial space.
In order to realize the efficient coupling of the multiscale phonon transport, the phonon scattering and advection are coupled together in the present scheme on the reconstruction of the distribution function at the cell interface.
The Newtonian method is adopted to solve the nonlinear scattering term for the update of the temperature at both the cell center and interface.
In addition, the energy at the cell center is updated by a macroscopic equation instead of taking the moment of the distribution function, which enhances the numerical conservation.
Numerical results of the cross-plane heat transfer prove that the present scheme can describe the multiscale heat transfer phenomena accurately with arbitrary temperature difference in a wide range.
In the diffusive regime, even if the time step is larger than the relaxation time, the present scheme can capture the transient thermal transport process accurately.
Compared to that under small temperature differences, as the temperature difference increases, the variation of the temperature distribution behaves quite differently and the average temperature in the domain increases in the ballistic regime but decreases in the diffusive regime.

\end{abstract}
\maketitle

\section{Introduction}

Over the past two decades, with the development and application of the highly integrated devices and materials, thermal management, including the heat guide, dissipation, overheating, hot spot etc, becomes a huge challenge~\cite{cahill2014nanoscale,cahill2003nanoscale,Majumdar98MET,ju1999microscale}.
To solve these problems, it is necessary to understand the thermal mechanism in these devices under various temperature ranges or length scales.
Actually, a number of experimental, theoretical and numerical studies~\cite{cahill2014nanoscale,cahill2003nanoscale,toberer2012advances,Minnich15advances,toberer2012advances} have been made to investigate the multiscale heat transfer phenomena in wide temperature and length ranges.
The phonon Boltzmann transport equation (BTE)~\cite{zhangZm07HeatTransfer,ZimanJM60phonons,ChenG05Oxford,srivastava1990physics} is one of the most widely used theories to study the heat transfer problems from tens of nanometers to hundreds of microns.

In order to solve the phonon BTE, some assumptions are adopted to simplify the computation.
For example, the isotropic wave vector space and the single-mode relaxation time approximation~\cite{srivastava1990physics,ZimanJM60phonons,ChenG05Oxford} are used to simplify the anisotropic phonon dispersion and the full phonon scattering kernel, respectively.
In most previous works, the temperature difference is assumed to be small enough~\cite{hua2015semi,LUO2017970} such that the relation between the temperature and the equilibrium state can be linearized~\cite{MurthyJY12HybridFBTE,narumanchi2004submicron} and the relaxation time can be treated as spatial independent.

Based on these assumptions, the size effects~\cite{xu_length-dependent_2014,PhysRevB.91.245423,PhysRevB.93.235423,chavez-angel_reduction_2014} with small temperature difference have been widely studied, but to the best of our knowledge, few studies have been devoted to a comprehensive understanding of the multiscale effects caused by large temperature difference.
Since the phonon relaxation time or the phonon mean free path depends on temperature~\cite{holland1963analysis,armstrong1985n}, the thermal conductivity may vary greatly for large temperature difference~\cite{holland1963analysis,terris2009modeling,toberer2012advances}.
In addition, different scattering mechanisms dominates the thermal conduction in different temperature ranges~\cite{MEHRA201892,narumanchi2004submicron,holland1963analysis,ChenG05Oxford,ZimanJM60phonons}.
For example, in extremely low temperature, the boundary scattering dominates the phonon transport.
Therefore, it is desired to study the multiscale heat transfer with large temperature differences.

For large temperature difference, the relationship between the temperature and the equilibrium state could not be linearized.
In addition, under this circumstance the phonon BTE is essentially multiscale, not only in the frequency space but also in the spatial space.
Actually, even with a given phonon frequency and polarization, the Knudsen number ($\text{Kn}$)~\cite{laurendeau2005statistical,ChenG05Oxford} may change significantly when the temperature difference is large.

Although it is a great challenge to solve the phonon BTE as the temperature difference is large, some numerical methods have been developed, such as the Monte Carlo method~\cite{MazumderS01MC,mittal2010monte,Lacroix05,randrianalisoa2008monte,HEATGENERATION11,Minnich15advances}, discrete ordinate method (DOM)~\cite{ChaiJC93RayEffect,SyedAA14LargeScale,MurthyJY05Review} and the hybrid Ballistic-Diffusive method~\cite{Pareekshith16BallisticDiffusive}.
The DOM is one of the most widely used deterministic methods, which discretizes the whole wave vector space into small pieces.
Due to the decoupling of the phonon scattering and advection, it usually suffers from large numerical dissipations in the diffusive regime.
Apart from deterministic methods, the statistic Monte Carlo method has also been widely employed in the study of phonon transport.
The main drawback of the Monte Carlo method is the restriction on the time step and cell size, namely, the time step and cell size have to be smaller than the relaxation time and mean free path.
For small Knudsen numbers, the method converges very slowly and suffers from large statistics errors.
In a word, few of the numerical methods mentioned above can predict accurate solutions for problems with large range of length and temperature variations.

Recently, a discrete unified gas kinetic scheme (DUGKS), which was firstly proposed for gas dynamics~\cite{GuoZl13DUGKS,GuoZl15DUGKS}, was developed to solve the multiscale heat transfer problem~\cite{GuoZl16DUGKS,LUO2017970}.
Different from the Monte Carlo method and the DOM, the phonon scattering and advection are coupled together in DUGKS by solving the phonon BTE which makes the cell size and time step not be restricted by the phonon mean free path and relaxation time.
The DUGKS can capture the phonon transport physics at different scales automatically with excellent numerical accuracy and stability.
Some progress based on the DUGKS has been made for multiscale heat transfer problem~\cite{GuoZl16DUGKS,LUO2017970}, but the thermal transport phenomena considered are limited to small temperature difference.

In this study, the DUGKS based on the non-gray phonon model~\cite{MurthyJY05Review,narumanchi2005comparison} with arbitrary temperature difference is developed.
The rest of this article is organized as follows. In Sec. 2, the phonon non-gray model is introduced, and in Sec. 3, the DUGKS with arbitrary temperature difference is provided in detail; in Sec. 4, the performances of the present scheme are validated by several tests, and the effects caused by temperature difference on thermal conduction with different temperature difference and length scales are studied; finally, a conclusion is drawn in the last section.

\section{Phonon Boltzmann transport equation}

The phonon Boltzmann transport equation (BTE) is a kinetic model to describe the phonon transport phenomena.
Under the single-mode relaxation time approximation~\cite{srivastava1990physics,ZimanJM60phonons,ChenG05Oxford}, it can be written as
\begin{equation}
\frac{\partial f}{\partial t}+ \bm v \cdot \nabla f  = \frac { f_0(T_{\text{loc}})-f }{\tau(T)},
\label{eq:BTE}
\end{equation}
where $f$ is the distribution function of phonons; $\bm v$ is the group velocity, and $\tau$ is the effective relaxation time which is a combination of all scattering processes~\cite{narumanchi2004submicron,holland1963analysis,armstrong1985n}.
The phonon equilibrium distribution $f^{eq}$ follows the Bose-Einstein distribution,
\begin{equation}
f^{eq}(\omega,T)=\frac {1}{\exp(\hbar\omega/k_{B}T)-1},
\label{eq:BoseE}
\end{equation}
where $\hbar$ is the Planck's constant divided by $2\pi$, $k_B$ is the Boltzmann constant, $\omega$ is the frequency and $T$ is the temperature.
$f_0$ is the equilibrium distribution function at the local pseudo-temperature $T_{\text{loc}}$~\cite{RAN2018616,MurthyJY05Review,ChenG05Oxford}, i.e., $f_0(T_{\text{loc}})=f^{eq}(T_{\text{loc}})$, which is introduced to ensure the energy conservation of the scattering term.
The detailed discussion of the local pseudo-temperature and temperature will be shown later.

The group velocity $\bm v= \nabla_{\bm{K}} {\omega} $ can be obtained through the phonon dispersion relation~\cite{brockhouse1959lattice,pop2004analytic}, where $\bm{K}$ is the wave vector.
We assume that the wave vector space is isotropic and $\bm{v}=|\bm{v}|\bm{s}$, where $\bm{s}=(\cos \theta, \sin \theta \cos \varphi, \sin \theta \sin \varphi)$ is the unit direction vector with $\theta$ the polar angle and $\varphi$ the azimuthal angle, respectively.
The distribution function $f=f( \bm x, \omega, p, \bm s, t)$ is related to the space position $\bm x$, frequency $\omega$, polarization $p$, $\bm s$ and time $t$.
Usually, the relaxation time $\tau=\tau(\omega, p, T)$ is a function of the frequency, polarization and temperature~\cite{holland1963analysis,armstrong1985n,terris2009modeling,Nanoletter11Hop}.
Therefore, the phonon mean free path $\lambda=|\bm{v}|\tau$ not only depends on the frequency and polarization, but also the temperature.
When the temperature difference is large, the system is essentially multiscale~\cite{Pareekshith16BallisticDiffusive} even for a given frequency and polarization.

The total energy $U$ and heat flux $\bm q$ can be obtained by taking the moments of the phonon distribution function over the whole wave vector space~\cite{ChenG05Oxford},
\begin{equation}
U =\sum_{p}\int_{\omega_{min,p}}^{\omega_{max,p}} \int_{4\pi} \hbar \omega f D(\omega, p)/4{\pi} d{\Omega}d{\omega},
\label{eq:U}
\end{equation}
\begin{equation}
\bm{q} =\sum_{p}\int_{\omega_{min,p}}^{\omega_{max,p}} \int_{4\pi} \bm v  \hbar \omega f D(\omega, p)/4{\pi} d{\Omega}d{\omega},
\label{eq:heatflux}
\end{equation}
where $D(\omega,p)$ is the phonon density of states~\cite{MazumderS01MC,ChenG05Oxford}, $\Omega$ is the solid angle, $\omega_{min,p}$ and $\omega_{max,p}$ are the minimum and maximum frequencies for a given phonon polarization branch $p$.
The temperature $T$ and the local pseudo-temperature $T_{\text{loc}}$ can be determined through the following constraints~\cite{RAN2018616,ChenG05Oxford},
\begin{equation}
0=\sum_{p}\int_{\omega_{min,p}}^{\omega_{max,p}} \int_{4\pi} \hbar \omega [f-f^{eq}(T)] D(\omega, p)/4{\pi} d{\Omega}d{\omega},
\label{eq:T}
\end{equation}
\begin{equation}
0=\sum_{p}\int_{\omega_{min,p}}^{\omega_{max,p}} \int_{4\pi} \hbar \omega \frac{ f-f_{0}(T_{\text{loc}})} {\tau(T)}  D(\omega, p)/4{\pi} d{\Omega}d{\omega}.
\label{eq:pseudoT}
\end{equation}

\section{DUGKS}

In this section, the DUGKS will be presented in detail.
In this method, the rectangle rule and the Gauss-Legendre~\cite{Abramovitch65Math,GuoZl16DUGKS} rule are used for the discretization of the frequency space and the solid angle space, respectively, and the trapezoidal quadrature is used for the time integration of the scattering term, while the mid-point rule is used for the flux term.
Equation~\eqref{eq:BTE} in integral form over a control volume can be written as follows,
\begin{equation}
f_{i,\omega, p,\alpha}^{n+1}-f_{i,\omega, p,\alpha}^{n} + \frac{\Delta t}{V_i} \sum_{j \in N(i)} \left( |\bm{v}|_{\omega,p} \bm{s}_{\alpha} \cdot \mathbf{n}_{ij} f_{ij,\omega, p,\alpha}^{n+1/2} S_{ij} \right)  =\frac{\Delta t}{2}\left( \frac{f_{0,i,\omega, p}^{n+1} -f_{i,\omega, p,\alpha}^{n+1} }{\tau_{i,\omega, p}^{n+1}}  +\frac{f_{0,i,\omega, p}^{n} -f_{i,\omega, p,\alpha}^{n} }{\tau_{i,\omega, p}^{n}}  \right),
\label{eq:dBTE}
\end{equation}
where $V_i$ is the volume of the cell $i$, $N(i)$ denotes the sets of neighbor cells of cell $i$, $ij$ denotes the interface between cell $i$ and cell $j$, $S_{ij}$ is the area of the  interface $ij$, $\mathbf{n}_{ij}$ is the normal unit vector of the interface $ij$ directing from cell $i$ to cell $j$; $\omega,~p,~\alpha$ are the index of the discretized phonon frequency, polarization and unit direction vector; $\Delta t$ is the time step from time $t_n$ to $t_{n+1}=t_{n}+ \Delta t$.
The formulation~\eqref{eq:dBTE} is implicit due to the scattering term at $t_{n+1}$.
In order to remove the implicitness, two new distribution functions are introduced and defined as
\begin{equation}
\tilde{f}=f-\frac{\Delta t}{2\tau}(f_0-f),
\label{eq:I1}
\end{equation}
\begin{equation}
\tilde{f}^{+}=f+\frac{\Delta t}{2\tau}(f_0-f).
\label{eq:I2}
\end{equation}
Then Eq.~\eqref{eq:dBTE} can be expressed as
\begin{equation}
\tilde{f}_{i,\omega, p,\alpha}^{n+1}-\tilde{f}_{i,\omega, p,\alpha}^{+,n}+ \frac{\Delta {t}}{V_i} \sum_{j \in N(i)} \left( |\bm{v}|_{\omega,p} \bm{s}_{\alpha} \cdot \mathbf{n}_{ij} f_{ij,\omega, p,\alpha}^{n+1/2} S_{ij} \right)  =0.
\label{eq:dBTE2}
\end{equation}

\begin{figure}
 \centering
 \includegraphics[width=0.80\textwidth]{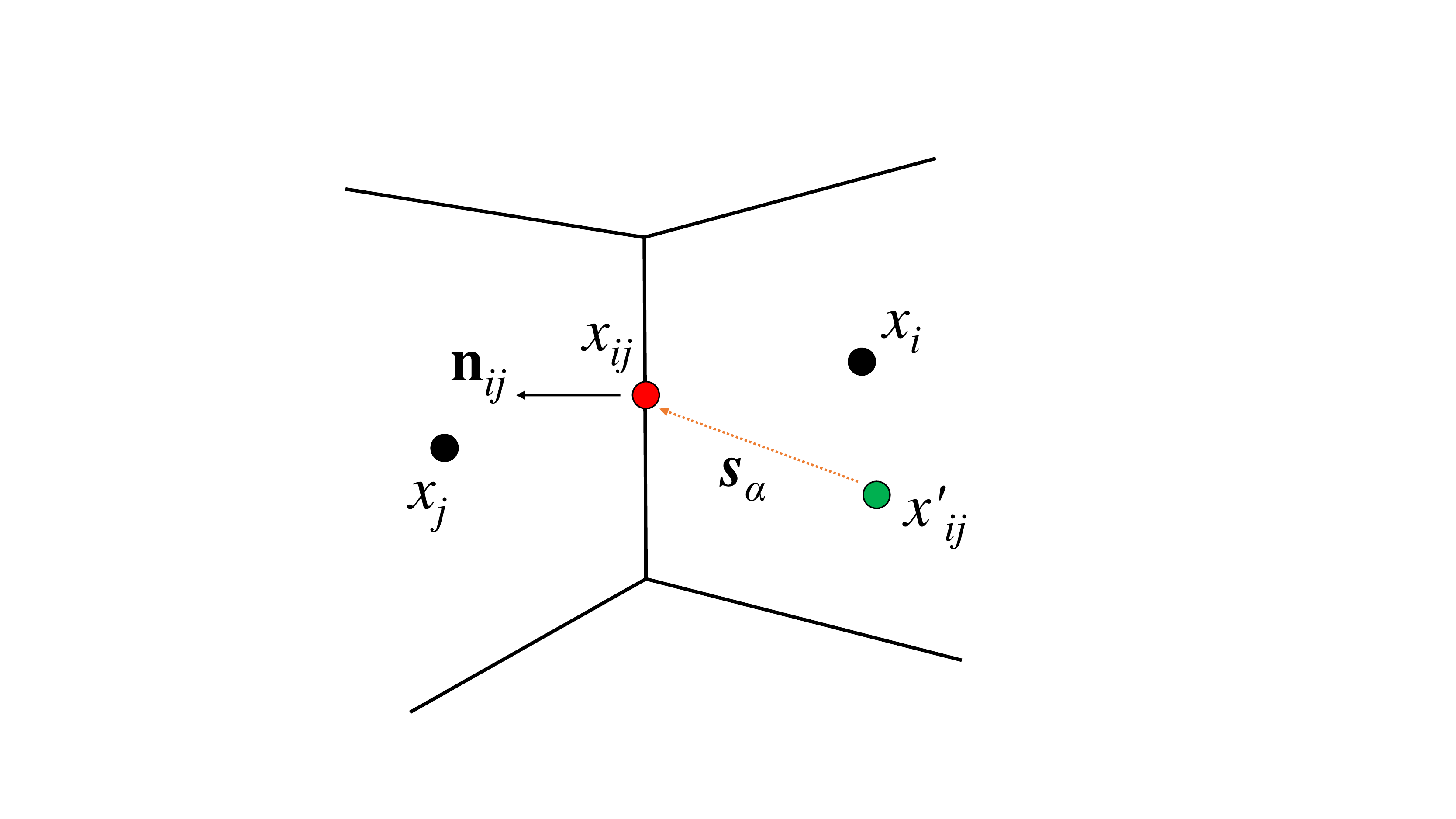}~~
 \caption{Schematic of phonon transport between two neighboring cells.}
 \label{dugkstransport}
\end{figure}

Different from direct numerical interpolation used in the DOM~\cite{ChaiJC93RayEffect,SyedAA14LargeScale}, in the DUGKS, the phonon BTE is employed on the reconstruction of the distribution function at the cell interface.
This is achieved by integrating Eq.~\eqref{eq:BTE} from time $t_n$ to $t_{n+1/2}=t_{n}+ \Delta t/2$ along the characteristic line with the end point $\bm{x}_{ij}$ locating at the center of the cell interface $ij$ between cell $i$ and cell $j$, i.e., (see~\cref{dugkstransport})
\begin{equation}
f_{\omega,p,\alpha}^{n+1/2}(\bm{x}_{ij})-f_{\omega,p,\alpha}^{n}(\bm{x}_{ij}')
=\frac{\Delta t}{4} \left[  \left.\   \left( \frac{f_{0,\omega,p,\alpha}^{n+1/2} -f_{\omega,p,\alpha}^{n+1/2} }{\tau_{\omega, p}^{n+1/2} } \right)  \right|_{\bm{x}=\bm{x}_{ij}}    +  \left.\  \left(  \frac{f_{0,\omega,p,\alpha}^{n} -f_{\omega,p,\alpha}^{n} }{\tau_{\omega, p}^{n}}  \right) \right|_{\bm{x}=\bm{x}_{ij}'}  \right],
\label{eq:fBTE}
\end{equation}
where $\bm{x}_{ij}'= \bm{x}_{ij}- |\bm{v}|_{\omega,p} \bm{s}_{\alpha}\Delta t/{2}$.
Equation~\eqref{eq:fBTE} can be reformulated as follows,
\begin{equation}
\bar{f}_{\omega,p,\alpha}^{n+1/2}(\bm{x}_{ij})-\bar{f}_{\omega,p,\alpha}^{+,n}(\bm{x}_{ij}') =0,
\label{eq:fBTE2}
\end{equation}
where
\begin{equation}
\bar{f}=f-\frac{\Delta t}{4\tau}(f_0-f),
\label{eq:I3}
\end{equation}
\begin{equation}
\bar{f}^{+}=f+\frac{\Delta t}{4\tau}(f_0-f).
\label{eq:I4}
\end{equation}
$\bar{f}_{\omega,p,\alpha}^{+,n}(\bm{x}_{ij}')$ is reconstructed as
\begin{equation}
\bar{f}_{\omega,p,\alpha}^{+,n}(\bm{x}_{ij}')=\bar{f}_{\omega,p,\alpha}^{+,n}(\bm{x}_{c}) + (\bm{x}_{ij}'-\bm{x}_{c}) \bm{\sigma}_{c},
\label{eq:slope}
\end{equation}
where $\bm{\sigma}_{c}$ is the spatial gradient of the distribution function $\bar{f}_{\omega,p,\alpha}^{+,n}(\bm{x}_{c})$  in the cell $c$.
If $\bm{s}_{\alpha} \cdot \bm{n}_{ij} >0$, $c=i$; else $c=j$.
The vanleer limiter~\cite{SwebyPK84Fluxlimiter} is adopted to determine the gradient to ensure the numerical stability and accuracy.

Combining Eqs.~\eqref{eq:fBTE2},~\eqref{eq:I4} and~\eqref{eq:slope}, the new distribution function $\bar{f}_{\omega,p,\alpha}^{n+1/2}(\bm{x}_{ij})$ at the cell interface at time $t_{n+1/2}$ can be obtained.
Based on Eqs.~\eqref{eq:I3} and~\eqref{eq:T}, together with the energy conservation of the scattering term, the temperature at the cell interface $T_{ij}$ at time $t_{n+1/2}$ can be obtained, i.e.,
\begin{equation}
0=\sum_{p}\int_{\omega_{min,p}}^{\omega_{max,p}} \int_{4\pi} \hbar \omega D/4\pi \left[ \bar{f}^{n+1/2} (\bm{x}_{ij})-f^{eq}(T_{ij}^{n+1/2}) \right] d{\Omega}d{\omega}.
\label{eq:fT}
\end{equation}
Similarly, the pseudo-temperature at the cell interface $T_{\text{loc},ij}$ at time $t_{n+1/2}$ can also be obtained based on Eqs.~\eqref{eq:I3} and~\eqref{eq:pseudoT}, i.e.,
\begin{equation}
0=\sum_{p}\int_{\omega_{min,p}}^{\omega_{max,p}} \int_{4\pi}  \hbar \omega D/4\pi \left[  \frac{\bar{f}^{n+1/2}(\bm{x}_{ij})-f_0(T_{\text{loc},ij}^{n+1/2}) }{4\tau(T_{ij}^{n+1/2})+\Delta t} \right] d{\Omega}d{\omega}.
\label{eq:fpseudoT}
\end{equation}
Consequently, $f_0(T_{\text{loc},ij}^{n+1/2})$ and $\tau(T_{ij}^{n+1/2})$ can be fully determined, and the distribution function $f_{ij}^{n+1/2}$ at the cell interface at time $t_{n+1/2}$ can be calculated based on Eq.~\eqref{eq:I3}.
Then, according to Eq.~\eqref{eq:dBTE2}, the distribution function $\tilde{f}_i^{n+1}$ at the cell center $i$ at time $t_{n+1}$ can be updated.

In order to mitigate the numerical quadrature errors, the total energy at the cell center is updated as follows,
\begin{equation}
\frac{U_i^{n+1}-U_i^{n}}{\Delta t}+  \frac{1}{V_i} \sum_{j \in N(i)} \bm{n}_{ij} \cdot \bm{q}_{ij}^{n+1/2} S_{ij}=0,
\label{eq:macro}
\end{equation}
where $\bm{q}_{ij}^{n+1/2}$ is obtained from Eq.~\eqref{eq:heatflux}.
By this method, the numerical errors introduced by the numerical quadrature of the heat flux can be mitigated and numerical conservation~\cite{LIU2018313} can be enhanced.
Then according to Eqs.~\eqref{eq:U}and~\eqref{eq:T}, the temperature $T_{i}^{n+1}$ can be updated by following equation,
\begin{equation}
U_i^{n+1}=\sum_{p}\int_{\omega_{min,p}}^{\omega_{max,p}} \int_{4\pi} \hbar\omega D/4\pi  {f}^{eq}(T_{i}^{n+1})  d{\Omega}d{\omega},
\label{eq:Tnext}
\end{equation}
The pseudo-temperature $T^{n+1}_{\text{loc},i}$ can be obtained similarly,
\begin{equation}
0=\sum_{p}\int_{\omega_{min,p}}^{\omega_{max,p}} \int_{4\pi}  \hbar \omega D/4\pi \left[  \frac{\tilde{f}_{i}^{n+1}-f_0(T_{\text{loc},i}^{n+1}) }{2\tau(T_{i}^{n+1})+\Delta t} \right] d{\Omega}d{\omega}.
\label{eq:pseudoTnext}
\end{equation}
The heat flux at the cell center can be computed from,
\begin{equation}
\bm{q} =\sum_{p}\int_{\omega_{min,p}}^{\omega_{max,p}} \int_{4\pi} \bm v  \hbar \omega \frac{2\tau}{2\tau+\Delta t}  \tilde{f} D(\omega, p)/4{\pi} d{\Omega}d{\omega},
\label{eq:heatflux2}
\end{equation}
where the relation between $f$ and $\tilde{f}$, i.e., Eq.~\eqref{eq:I1}, has been considered.

Noting that the Newtonian iteration method is used for the solution of the temperature and the pseudo-temperature at both the cell center and interface.
The new distribution functions $\tilde{f},~\tilde{f}^{+},~\bar{f},~\bar{f}^{+}$ are all related to the original distribution function $f$ and the equilibrium state $f_0$.
Some relations, which are used in the computation, are depicted as follows,
\begin{equation}
\tilde{f}^{+}=\frac{4}{3}\bar{f}^{+}-\frac{1}{3}\tilde{f},
\label{eq:relationa1}
\end{equation}
\begin{equation}
\bar{f}^{+}=\frac{4\tau-\Delta t}{4\tau+2\Delta t}\tilde{f}+ \frac{3\Delta t}{4\tau+2\Delta t} {f}_0.
\label{eq:relationa2}
\end{equation}
The time step of the DUGKS is determined by the Courant-Friedrichs-Lewy (CFL) condition, i.e.,
\begin{equation}
\Delta t=\eta \frac{\Delta x_{\text{min}}}{v_{\text{max}}},
\label{eq:cfl}
\end{equation}
where $\eta$ is the CFL number and $\eta \in (0,1)$, $\Delta x_{\text{min}}$ is the minimum cell size and $v_{\text{max}}$ is the maximum group velocity over all phonon frequencies and branches.
It can be shown that the DUGKS described above has the asymptotic preserving property~\cite{GuoZl13DUGKS,GuoZl15DUGKS}, which has been discussed in previous works~\cite{GuoZl16DUGKS}.
Consequently the time step could be larger than the relaxation time even in the diffusive regime.
The boundary conditions are similar to those discussed in the previous studies~\cite{GuoZl16DUGKS,LUO2017970,SyedAA14LargeScale}.

The main procedure of the present DUGKS from time $t_n$ to $t_{n+1}$ can be summarized as follows,
\begin{enumerate}
  \item calculate $\bar{f}_{\omega,p,\alpha}^{+,n}$ and $\tilde{f}_{\omega,p,\alpha}^{+,n}$ based on Eqs.~\eqref{eq:I4} and~\eqref{eq:I2}, respectively;
  \item calculate the spatial gradient and reconstruct the $\bar{f}_{\omega,p,\alpha}^{+,n}(\bm{x}_{ij}')$ based on Eq.~\eqref{eq:slope};
  \item calculate the $\bar{f}_{\omega,p,\alpha}^{n+1/2}(\bm{x}_{ij})$ based on Eq.~\eqref{eq:fBTE2};
  \item calculate the temperature and the local pseudotemperature at time $t_{n+1/2}$ on the cell interface based on Eqs.~\eqref{eq:fT} and~\eqref{eq:fpseudoT} in sequence;
  \item calculate the original distribution function at the cell interface ${f}_{\omega,p,\alpha}^{n+1/2}(\bm{x}_{ij})$ based on Eq.~\eqref{eq:I3}, then calculate the macroscopic heat flux across the cell interface $\bm{q}^{n+1/2}(\bm{x}_{ij})$ from the moment of the distribution function ${f}_{\omega,p,\alpha}^{n+1/2}(\bm{x}_{ij})$, i.e., Eq.~\eqref{eq:heatflux};
  \item update the distribution function $\tilde{f}_{\omega,p,\alpha}^{n+1}$ at the cell center based on Eq.~\eqref{eq:dBTE2};
  \item update the macroscopic distribution ($T$, $T_{\text{loc}}$, $\bm{q}$) at the cell center at the next time step based on Eqs.~\eqref{eq:macro},~\eqref{eq:Tnext},~\eqref{eq:pseudoTnext} and~\eqref{eq:heatflux2}, respectively.
\end{enumerate}

\section{Numerical results and discussion}

In this section, some numerical tests will be carried out to validate the present DUGKS with arbitrary temperature differences.
The effects on thermal conduction caused by different temperature difference will also be discussed.

\begin{figure}
 \centering
 \includegraphics[width=0.90\textwidth]{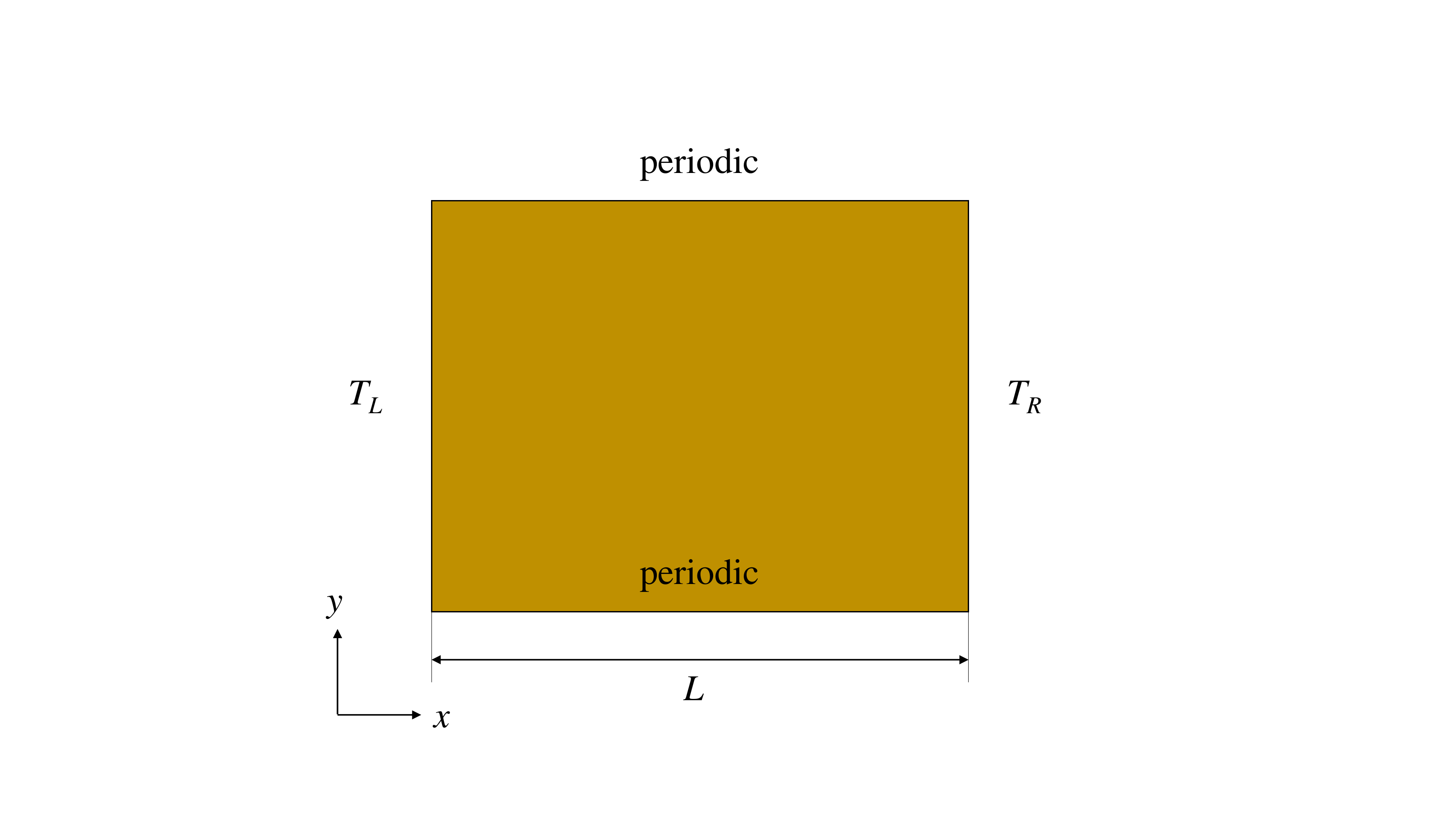}~~
 \caption{Schematic of the 2D square thin film.}
 \label{filmfig}
\end{figure}

The problem considered is the cross-plane heat transfer.
Considering a square 2D thin film of thickness $L$ (as depicted in~\cref{filmfig}), the top and bottom boundaries are set to be periodic.
The temperatures of the left and right boundaries are set to be $T_L=T_0+\Delta T/2$ and $T_R=T_0-\Delta T/2$, respectively, where $T_0= (T_L+T_R)/2$ and $\Delta T$ is the temperature difference, respectively.
The isothermal boundary conditions are implemented on the two boundaries.
A temperature ratio, $R=\Delta T/T_{0}$, is defined as the relative temperature difference.

In the tests, monocrystalline silicon~\cite{brockhouse1959lattice} is used as the material.
The phonon dispersion relation of the silicon in the [100] direction is chosen~\cite{holland1963analysis,brockhouse1959lattice}.
The optical phonon branches are not considered due to the little contribution to the thermal conduction.
The Pop's dispersion relation~\cite{pop2004analytic} and the Terris' method for the relaxation time~\cite{terris2009modeling} are employed in the simulations (see Appendix).
For each phonon branch, the wave vector $k \in [0,2\pi/a]$ is discretized equally into $N_B$ discrete bands, i.e., $k_b=(2\pi/a)(2b-1)/(2N_B)$, where $b$ is an index and $ 1 \leq b \leq N_B$, $a$ is the lattice constant and $a=5.43$\r{A} for silicon.
The corresponding discretized frequency can be obtained through the dispersion relations $\omega_b=\omega(k_b,p)$~\cite{pop2004analytic}.
For example, $\omega_{\text{min},p}=0,~\omega_{\text{max},p}=\omega(2\pi/a,p)$.
The cartesian grid is used to discretize the physical space, and $N_x,~N_y$ uniform cells are used in the $x$ and $y$ direction, respectively.
The $\cos \theta \in[-1,1]$ is discretized with $N_{\theta}$-point Gauss-Legendre quadrature, while the azimuthal angle $\varphi \in [0,\pi]$ (due to symmetry) is discretized with $N_{\varphi}$-point Gauss-Legendre quadrature.
Without special statements, we set $N_B=40$, $\eta=0.80$, and $N_y=4$, $N_{\varphi}=4$ due to the periodicity of the heat transfer in the $y$ direction.
For steady problem, the system is regarded as converged as $\epsilon<10^{-6}$, where
\begin{equation}
\epsilon=\frac{ \sqrt {\sum_{i}{(T_i^{n}-T_i^{n+1000})^2}}} { \sqrt { \sum_{i} (\Delta T)^2 } }.
\label{eq:epsilon}
\end{equation}

\begin{figure}
 \centering
 \subfloat[]{\includegraphics[width=0.45\textwidth]{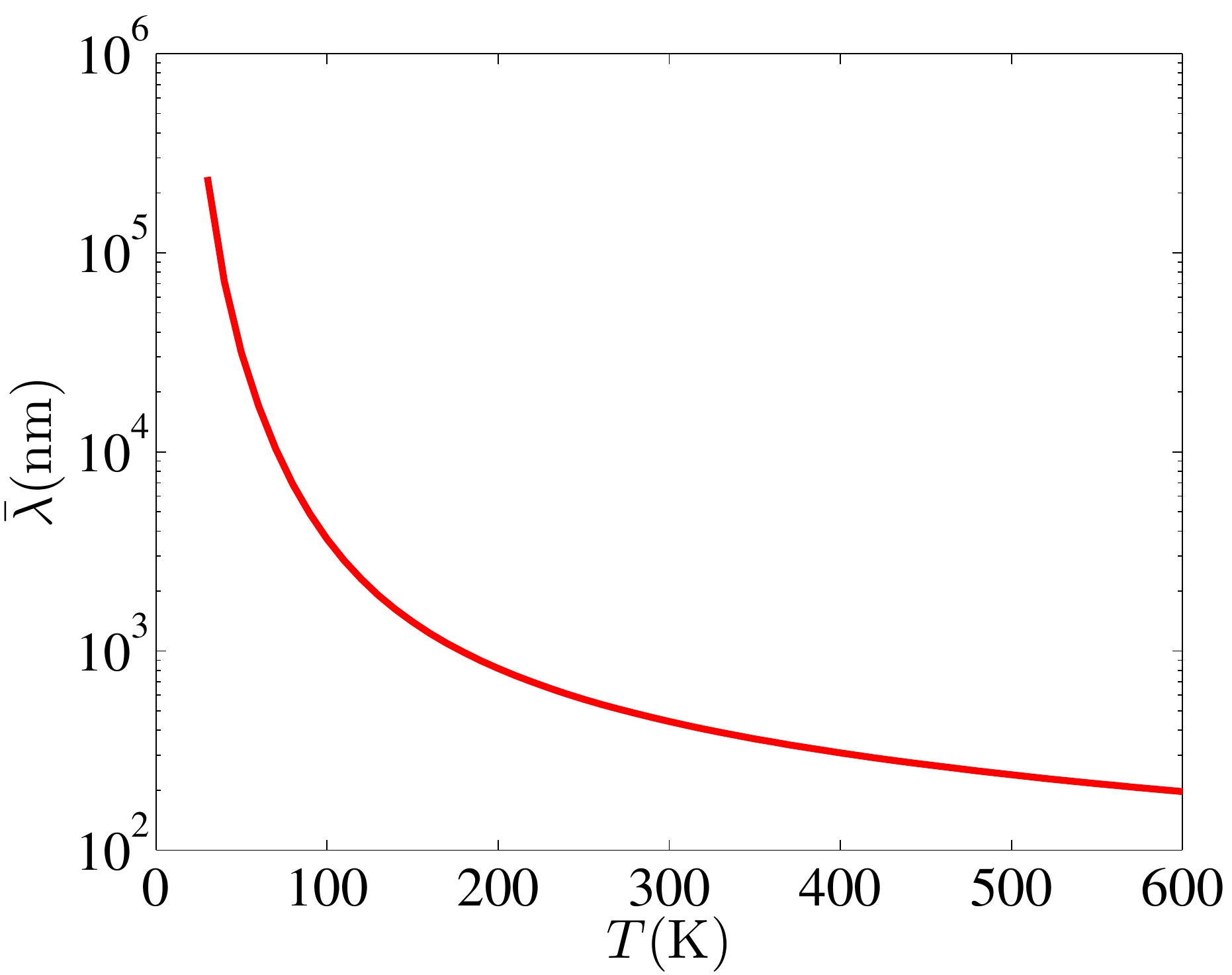}}~~
 \subfloat[]{\includegraphics[width=0.45\textwidth]{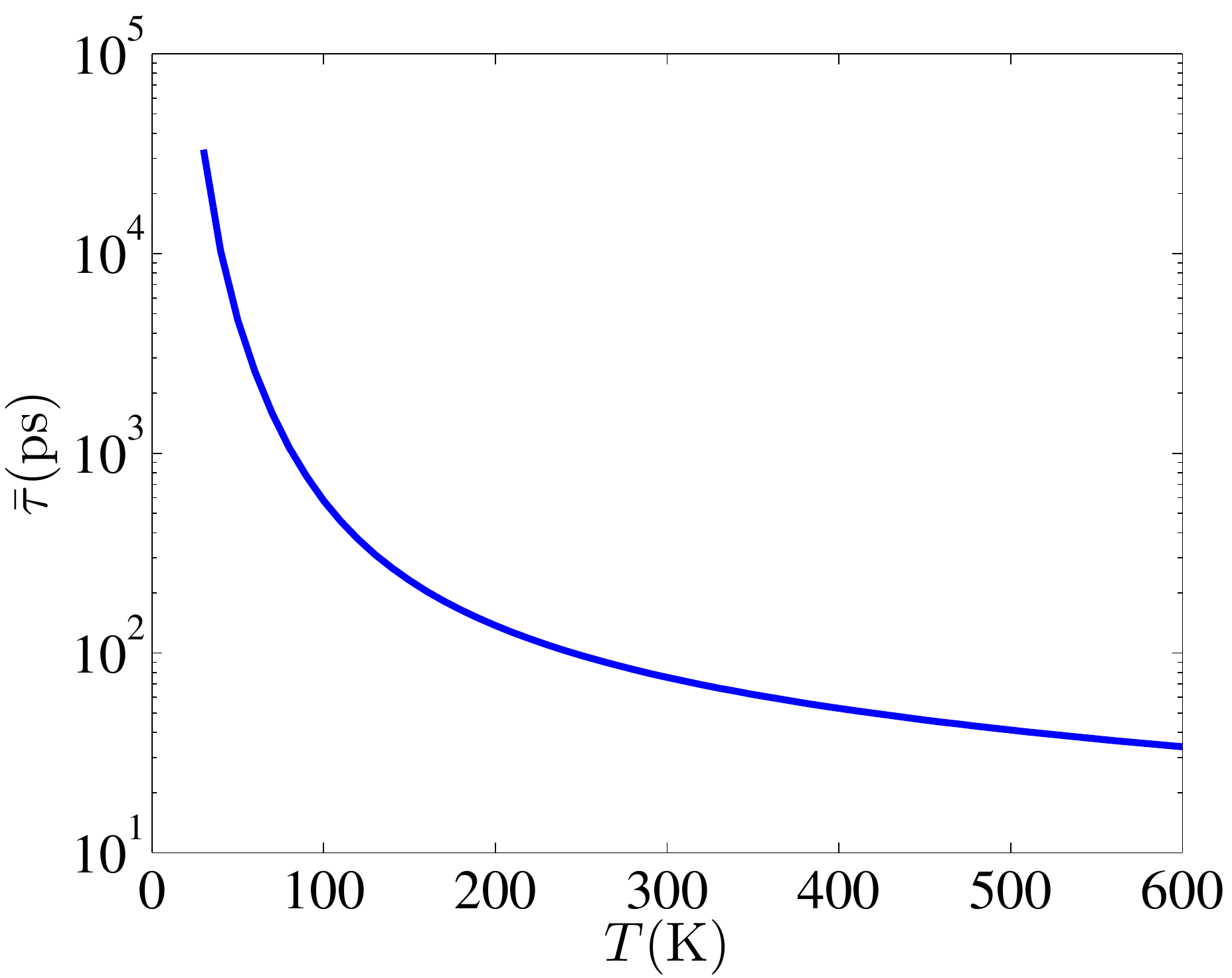}}~~
 \caption{ (a) Temperature-dependent average phonon mean free path (Eq.~\eqref{eq:mfp}), (b) Temperature-dependent average relaxation time (Eq.~\eqref{eq:avetau}).}
 \label{mfp}
\end{figure}

The phonon mean free path or relaxation time with different frequencies at room temperature may range from several nanometers to hundreds of microns, or from a few picoseconds to several nanoseconds.
For the sake of description, the average mean free path $\bar{\lambda}$ and average relaxation time $\bar{\tau}$ are introduced and defined as
\begin{equation}
\bar{\lambda} \left( T \right)=\left( \sum_{p}\int_{\omega_{min,p}}^{\omega_{max,p}} \hbar \omega D  \frac{d f^{eq}}{dT}  |\bm{v}| \tau d{\omega}  \right)  \left( \sum_{p}\int_{\omega_{min,p}}^{\omega_{max,p}} \hbar \omega D  \frac{d f^{eq}}{dT}  d{\omega}  \right)^{-1},
\label{eq:mfp}
\end{equation}
\begin{equation}
\bar{\tau} \left( T \right)=\left( \sum_{p}\int_{\omega_{min,p}}^{\omega_{max,p}} \hbar \omega D  \frac{d f^{eq}}{dT}  |\bm{v}| \tau d{\omega}  \right)  \left( \sum_{p}\int_{\omega_{min,p}}^{\omega_{max,p}} \hbar \omega D  |\bm{v}|  \frac{d f^{eq}}{dT}  d{\omega}  \right)^{-1}.
\label{eq:avetau}
\end{equation}
Then an average Knudsen number can be defined as
\begin{equation}
\overline{\text{Kn}} (T)=\bar{\lambda}/L.
\label{eq:kn}
\end{equation}
From the profiles shown in~\cref{mfp}, it can be observed that the average mean free path or relaxation time increases as $T$ decreases.
The heat transfer in different spatial domain may be in different regimes as the temperature difference is large.
As $T=300 \text{K}$, the corresponding average mean free path and relaxation time are $0.44\mu$m and $75$ps, respectively.

\subsection{Validation}

\begin{figure}
 \centering
 \includegraphics[width=0.45\textwidth]{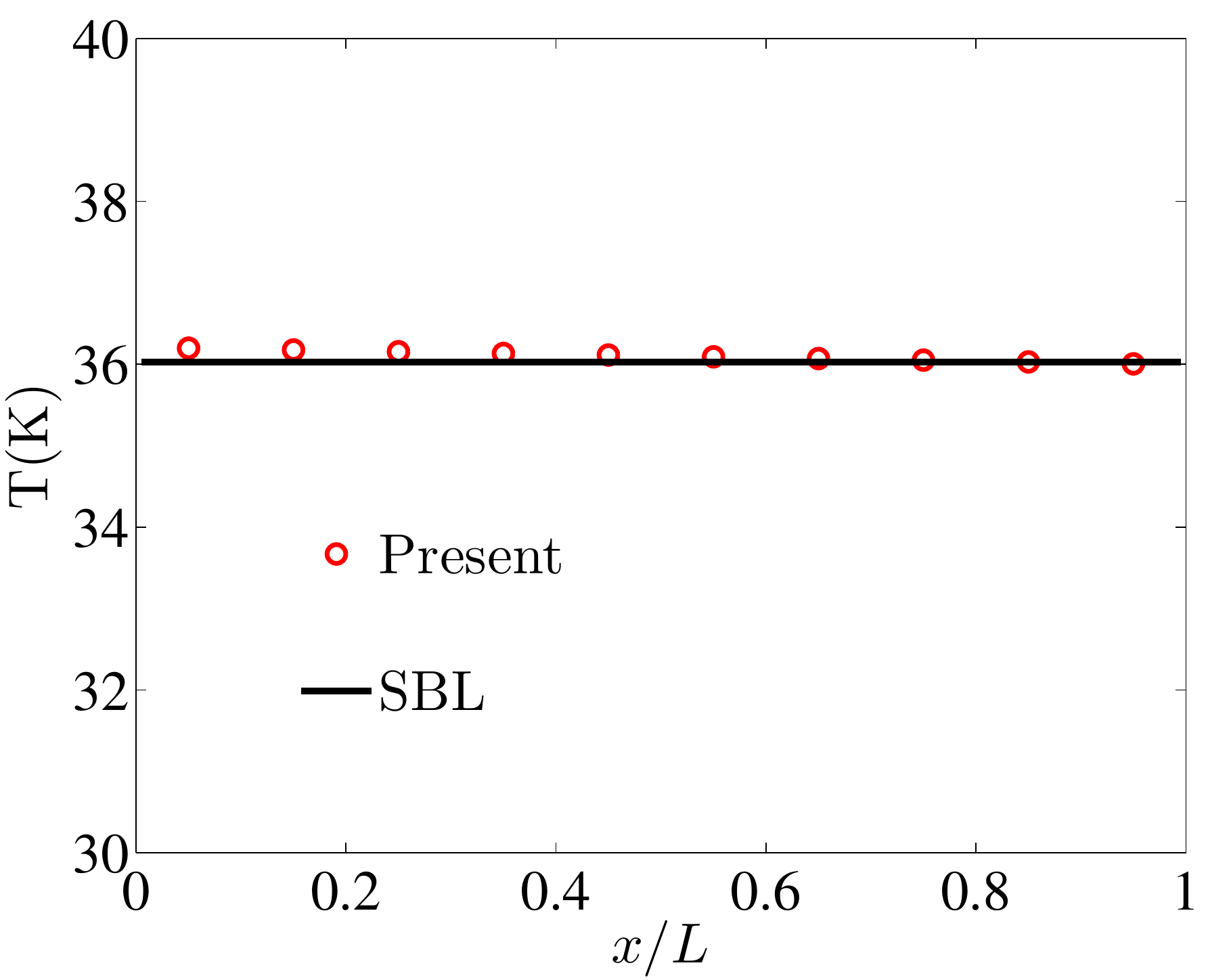}~~
 \caption{Temperature profiles with different thickness of the film. SBL means the analytical solution of the Stefan-Boltzmann law~\cite{MajumdarA93Film} in the ballistic limit.
 }
 \label{ballisticlimit}
\end{figure}
\begin{figure}
 \centering
 \subfloat[]{\includegraphics[width=0.45\textwidth]{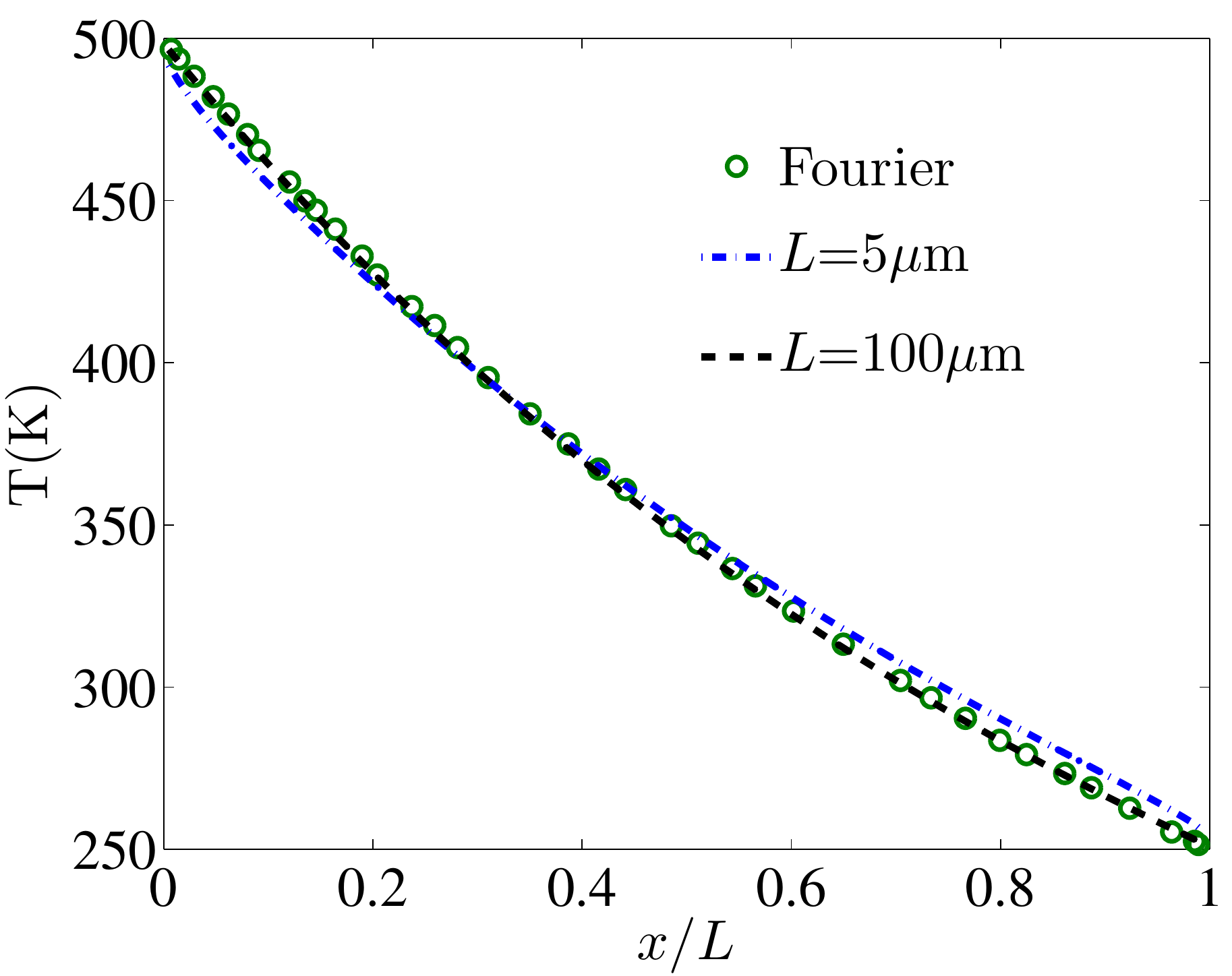}}~~
 \subfloat[]{\includegraphics[width=0.45\textwidth]{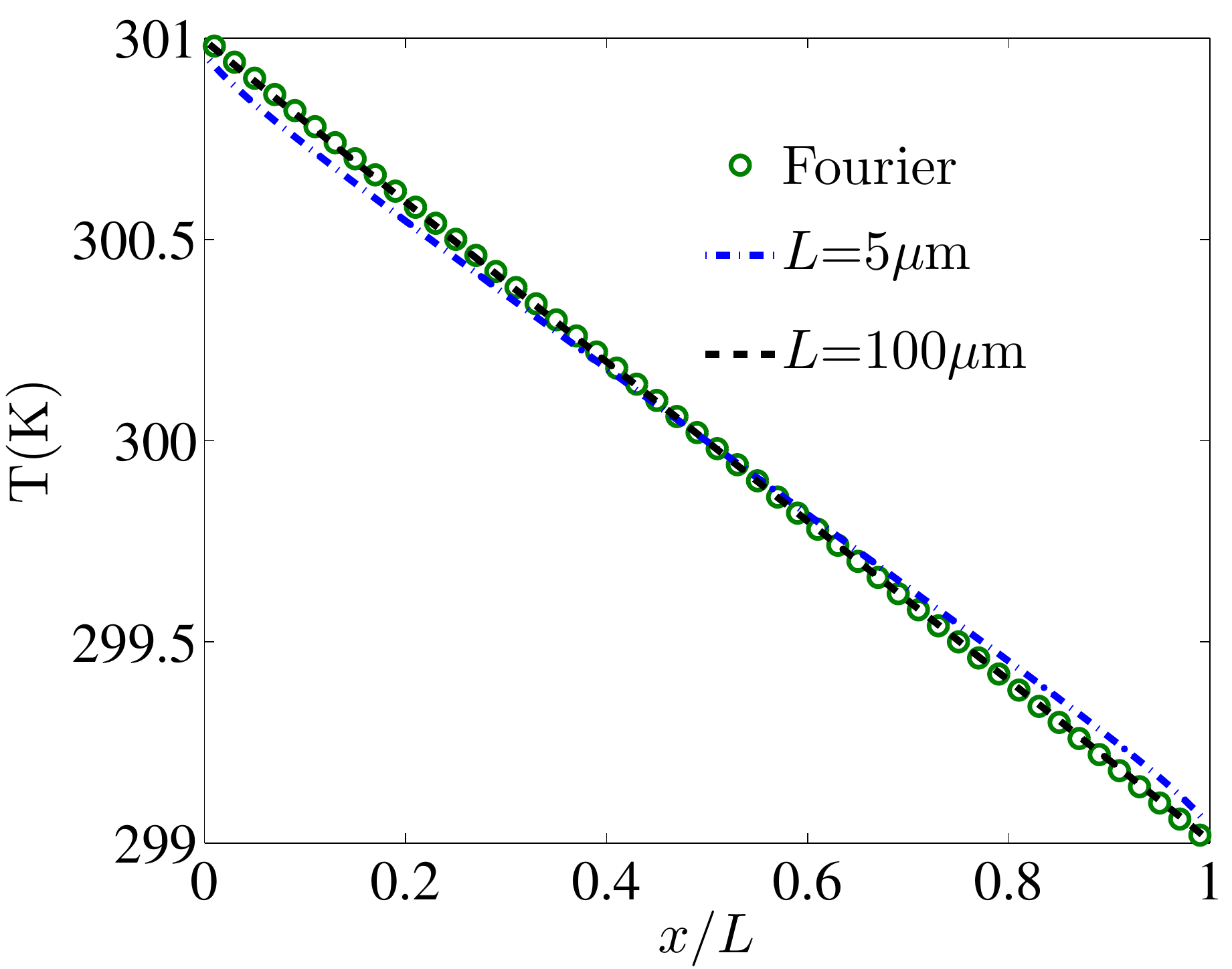}}~~
 \caption{Temperature profiles with different thickness of the film. Fourier solution is obtained from Ref~\cite{Lacroix05} with the temperature dependent thermal conductivity of silicon. (a) $T_L=500 \text{K}$, $T_R=250 \text{K}$, (b) $T_L=301 \text{K}$, $T_R=299 \text{K}$.
 }
 \label{diffusivefilm}
\end{figure}

\begin{figure}
 \centering
 \includegraphics[width=0.5\textwidth]{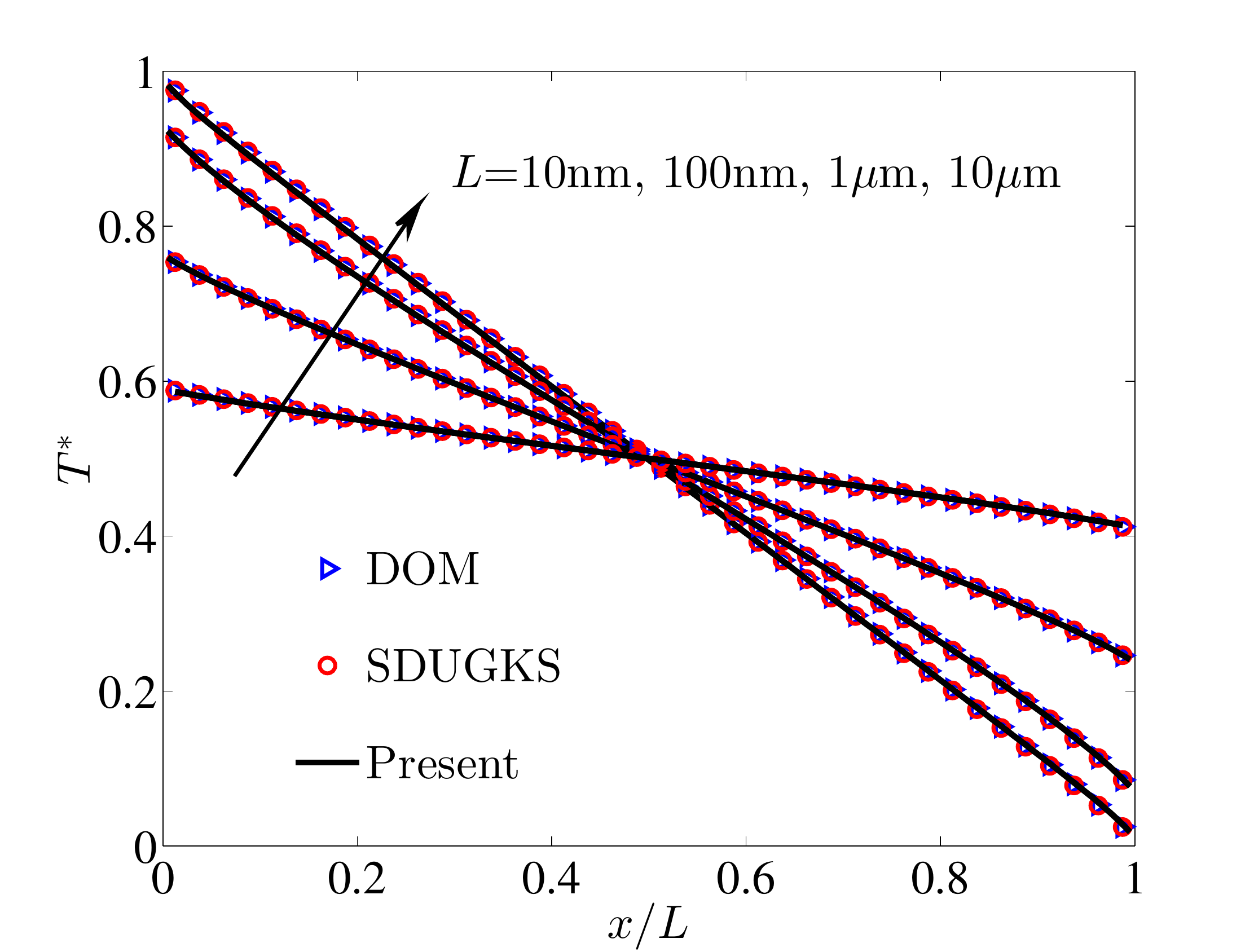}~~
 \caption{Non-dimensional temperature with different film thickness. $T_0=300\text{K}$, $T^{*}=(T-T_R)/(T_L-T_R)$. The blue triangle and the red circle are the data predicted by the implicit DOM and the DUGKS under the small temperature difference assumption (labeled as DOM and SDUGKS)~\cite{wangmr17callaway,LUO2017970}, respectively. The black solid line is the present results with $\Delta T/T_{0}=0.01$.
 }
 \label{smallTvalidation}
\end{figure}

The performance of the present scheme in both ballistic and diffusive regimes is first tested.
In the ballistic limit, phonon scattering is rare.
The phonon transport is governed by the Stefan-Boltzmann law~\cite{MajumdarA93Film} and the temperature in the whole domain is constant, namely $T^4=(T_{L}^{4}+T_{R}^{4})/2$.
To simulate this regime, we set $T_L=40~\text{K}$, $T_R=30~\text{K}$, $L=10 \text{nm}$, then $\overline{\text{Kn}}(T_L) \gg 10$.
$N_x=10$ cells and $N_{\theta}=100$ directions are used, which are enough to capture the highly non-equilibrium heat transfer.
The predicted temperature is shown in~\cref{ballisticlimit}.
The present numerical results are in good agreement with the analytical solution of the Stefan-Boltzmann law~\cite{MajumdarA93Film}.

With the increasing of the thickness of the film, the intrinsic scattering becomes important.
The thermal transport phenomena with $L = 5\mu$m and $100\mu$m are studied with the present scheme in two different temperature ranges, i.e., $T_L=500~{\text{K}}$, $T_R=250~{\text{K}}$ and $T_L=301~{\text{K}}$, $T_R=299~{\text{K}}$.
Then we have $\overline{\text{Kn}}(T_R) < 0.2 $, which indicates that the heat transfer is close to that in the diffusive regime.
We set $N_x  =100,~N_{\theta} =40$ and the results are shown in~\cref{diffusivefilm}.
The temperature distributions are compared with the Fourier solutions obtained from Ref~\cite{Lacroix05}.
From the profiles, it can be observed that the small temperature jump on the boundary still exists as $L=5$$\mu$m.
In other words, as $L=5$$\mu$m, the Fourier law is invalid.
As $L=100\mu$m, the results keep consistent with that predicted by the Fourier law.
It is also observed that as the temperature difference is large, the temperature profiles are nonlinear due to the temperature-dependent thermal conductivity~\cite{Lacroix05,holland1963analysis}.

Base on above results, it can be concluded that the present scheme can capture the thermal transport accurately in both ballistic and diffusive regimes.
Then the performance of the present scheme with small temperature difference is tested in transition regime.
We consider the case of $R=0.01$ and $T_0=300 \text{K}$ with different film thickness.
Figure~\ref{smallTvalidation} shows the distribution of the non-dimensional temperature.
It can be seen that the present results are consistent with the data obtained by the implicit DOM and the DUGKS under the small temperature difference assumption~\cite{wangmr17callaway,LUO2017970}.
These results confirm that for problems with small temperature difference, the present scheme can recover the multiscale phonon transport physics correctly.

\subsection{Transient heat transfer}

The transient heat transfer is studied in this subsection.
The temperature is initialized as,
\begin{equation}
T(x,t)=T_R,~t=0,~0<x<L.
\end{equation}
In the diffusive regime, the heat transfer can be described by the Fourier equation~\cite{d1986analysis,holman1989heat,MurthyJY05Review}, i.e.,
\begin{equation}
C(T)\frac{\partial T}{\partial t }= \nabla \cdot (K(T) \nabla T ),
\label{eq:fourier}
\end{equation}
where $C(T)$, $K(T)$ are the specific heat and thermal conductivity in the diffusive limit~\cite{MurthyJY05Review}, respectively.
The traditional finite volume method with enough discretized cells is used to solve Eq.~\eqref{eq:fourier}~\cite{d1986analysis,holman1989heat}.
As the temperature difference is small, we can assume $C$ and $K$ are constants and an analytical solution can be obtained~\cite{Lacroix05,holman1989heat},
\begin{equation}
\frac{T(x,t)-T(L,t) }{T(0,t)-T(L,t) } =
\text{erfc}  \left( \frac{x}{2 \sqrt{\beta t}}  \right)
 - \text{erfc} \left( \frac{ 2L-x }{ 2  \sqrt{\beta t}}  \right) + \text{erfc} \left( \frac{ 2L+x }{2  \sqrt{\beta t}}  \right),
 \label{eq:laplace}
\end{equation}
where $\beta =K/C$ is the thermal diffusivity.
As $T=300\text{K}$, $\beta = 1.48 \text{cm}^2 s^{-1}$ in our simulations.

\begin{figure}
 \centering
  \subfloat[]{\label{film41}\includegraphics[width=0.45\textwidth]{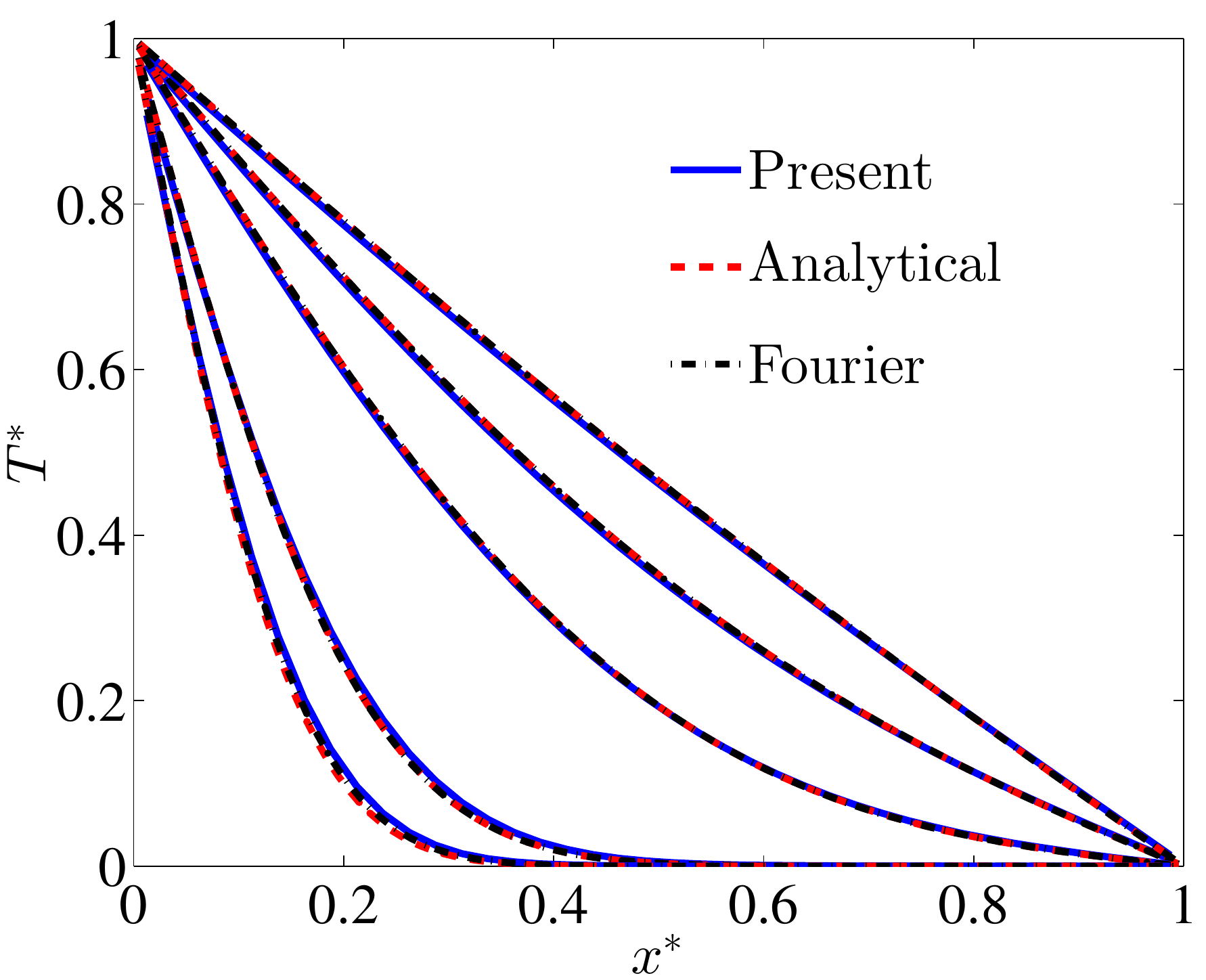}}~~
 \subfloat[]{\label{film42}\includegraphics[width=0.45\textwidth]{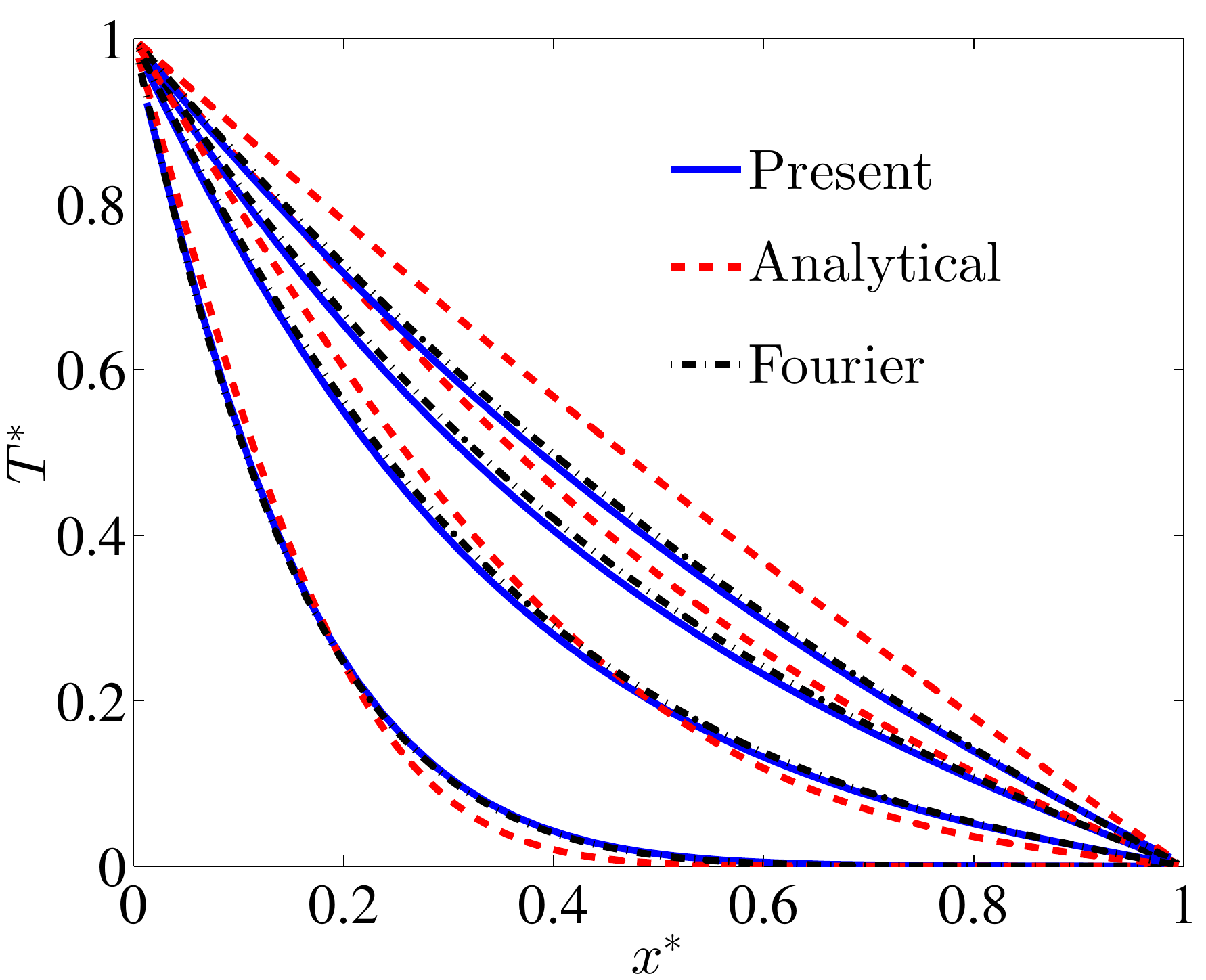}}~~ \\
 \caption{Non-dimensional temperature profiles in the diffusive regime, where $x^*=x/L$, $T^*=(T-T_R)/(T_L-T_R)$. $L=100$$\mu$m, $T_0=300\text{K}$. The label 'Fourier' is the solution of the Fourier law, and the solution based on the Eq.~\eqref{eq:laplace} is regarded as the analytical solution. The corresponding time is $t^*=2,~10,~20,~40$ from the left to right, respectively, where $t^*=t/t_0$, $t_0=500 \text{ns}$. (a) $\Delta T/T_0=0.01$, (b) $\Delta T/T_0=0.5$. }
 \label{transientT300L100umDT}
\end{figure}

The transient heat transfer in the diffusive regime is first tested.
We set $T_0=300\text{K}$, $L=100 $$\mu$m, $N_{x}=40$, $N_{\theta} =4 ,~\Delta t=100.0 \text{ps}$, which indicates $\overline{\text{Kn}}(T_0)=0.004$, $\Delta {x} / \bar{\lambda} = 5.7 $ and $\Delta t / \bar{\tau}= 1.3$.
Two temperature differences are considered, i.e., $\Delta T/T_0 =0.01$ or $0.5$.
The discretized cells and directions are enough to capture the heat transfer correctly.
The temperature profiles at different times are shown in~\cref{transientT300L100umDT}.
It can be observed that the results predicted by the present method are in excellent agreement with the data obtained by the Fourier equation no matter the temperature difference is small or large.
For the analytical solution of Eq.~\eqref{eq:laplace}, as the temperature difference is small (\cref{film41}), the results match well with the data predicted by other two methods.
But as the temperature difference is large (\cref{film42}), the numerical deviations are large, because it does not consider the change of the thermal diffusivity in the spatial space.

\begin{figure}
 \centering
  \subfloat[]{\label{film51}\includegraphics[width=0.45\textwidth]{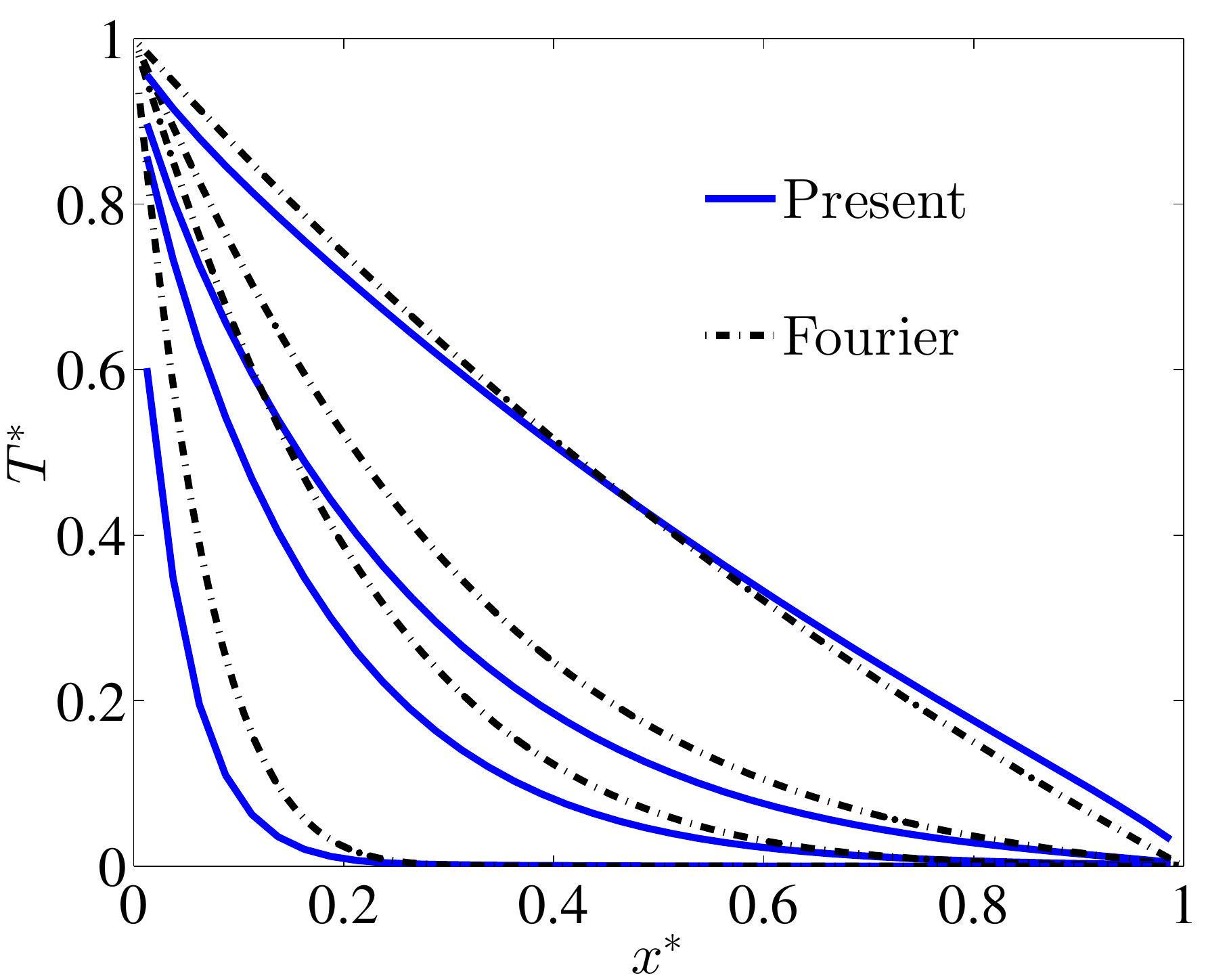}}~~
 \subfloat[]{\label{film52}\includegraphics[width=0.45\textwidth]{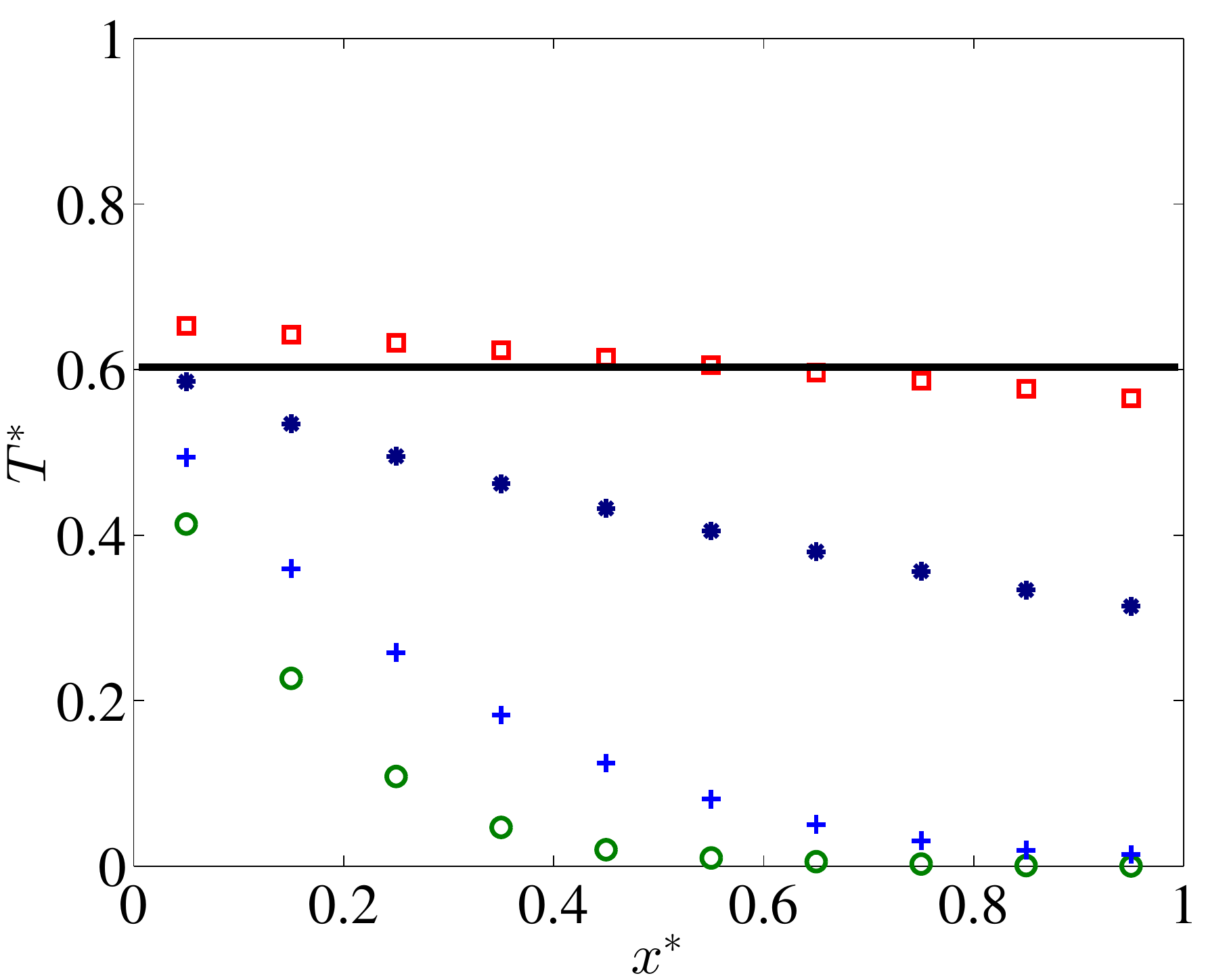}}~~\\
 \caption{Non-dimensional temperature profiles in different regimes, where $x^*=x/L$, $T^*=(T-T_R)/(T_L-T_R)$. (a) $L=5\mu$m, $\Delta T/T_{0}=0.5,~T_0=300 \text{K}$. From the left to right, $t=0.5 \text{ns},~5 \text{ns},~10 \text{ns},~100 \text{ns}$. The label 'Fourier' is the solution of the Fourier law. (b) $L=100 \text{nm}, ~T_L=40\text{K},~T_R=30\text{K}$. The scatters are the present results at different moments. From the left to right, $t=10 \text{ps},~20 \text{ps},~100 \text{ps},~5000 \text{ps}$. The black solid line is the analytical solution of the Stefan-Boltzmann law~\cite{MajumdarA93Film} in the ballistic limit.}
 \label{sizeeffects}
\end{figure}

Finally, the transient heat transfer with large temperature difference in non-diffusive regime is simulated.
We set $L=5 $$\mu$m, $T_0=300\text{K}$, $\Delta T/T_0 =0.5$, $N_{x}=40 $.
In order to capture the non-equilibrium effects accurately, more directions are used, for example, $N_{\theta} =16$.
The numerical results are shown in~\cref{film51}.
It can be observed that the temperature predicted by the present scheme is smaller than that obtained by the Fourier equation as $t<10$ns.
This is reasonable since as $L=5 $$\mu$m the size effects exist and reduce the thermal conductivity.
To study the transient heat transfer phenomena in the ballistic regime, we further simulate the case with $L=100 \text{nm}, ~T_L=40\text{K},~T_R=30\text{K}$.
Under this circumstance, $\overline{\text{Kn}}(T_L) \gg 10$ so that less cells and more directions can be used.
We set $N_{x}=10$, $N_{\theta}=100 $.
From the data shown in~\cref{film52}, it can be found that about $10-20 ~\text{ps}$ are needed to heat the cell closest to the cold boundary.
This is in agreement with the phonon group velocity calculated by the phonon dispersion curves of silicon.

This transient test shows that the present scheme can capture the unsteady heat transfer accurately.
In addition, it can be found that the time step and the cell size used in the diffusive regime can be larger than the average relaxation time and mean free path.

\subsection{Effects of temperature difference}

The effects of temperature difference at steady state are investigated in this subsection.
In order to capture the multiscale heat transfer physics, we set $N_{\theta}=100$ and change $N_x$ from $10$ to $100$ as $\overline{\text{Kn}}$ decreases.
Several cases with different temperature difference and film thickness are simulated.

\begin{figure}
 \centering
  \subfloat[]{\label{film16}\includegraphics[width=0.45\textwidth]{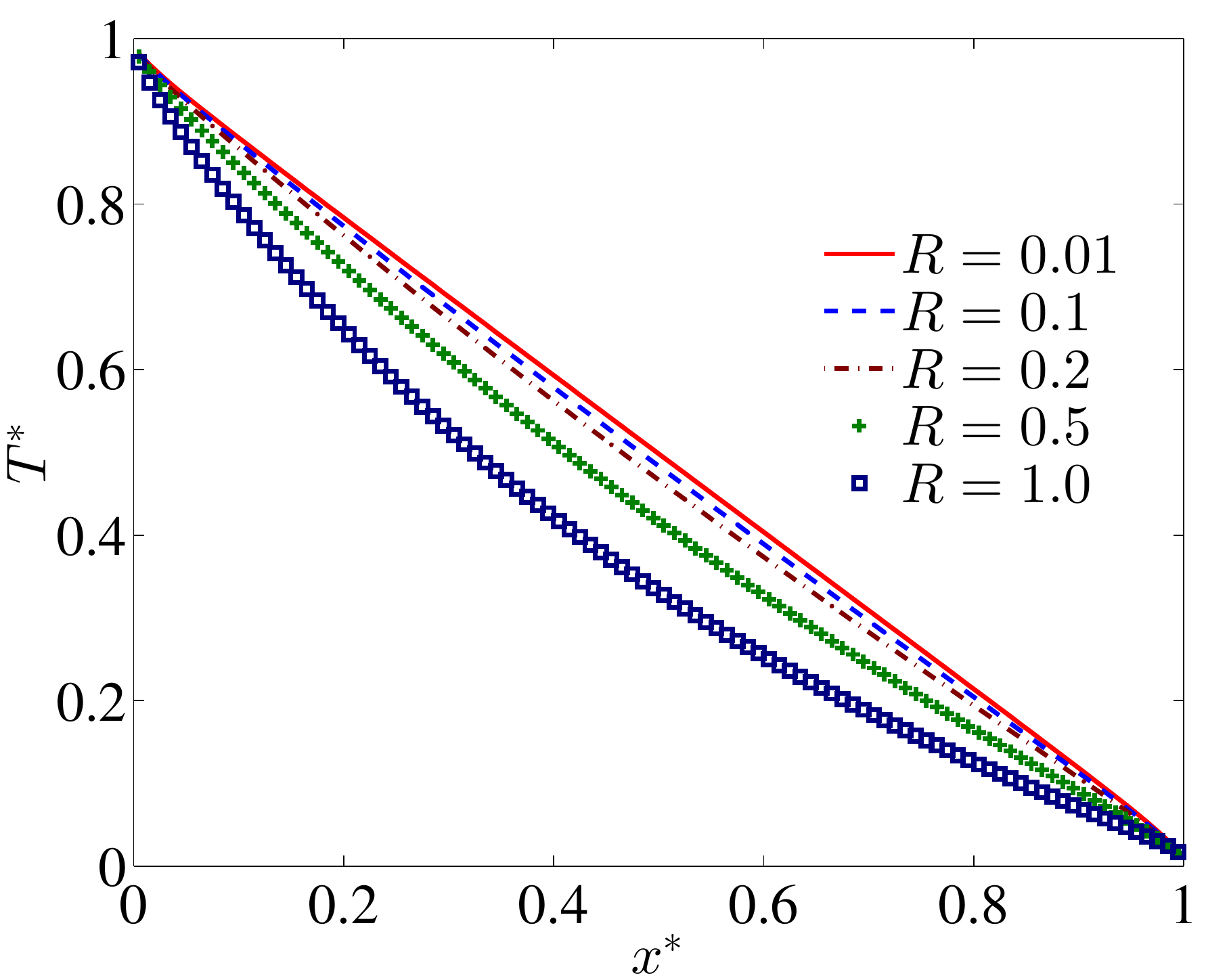}}~~
 \subfloat[]{\label{film19}\includegraphics[width=0.45\textwidth]{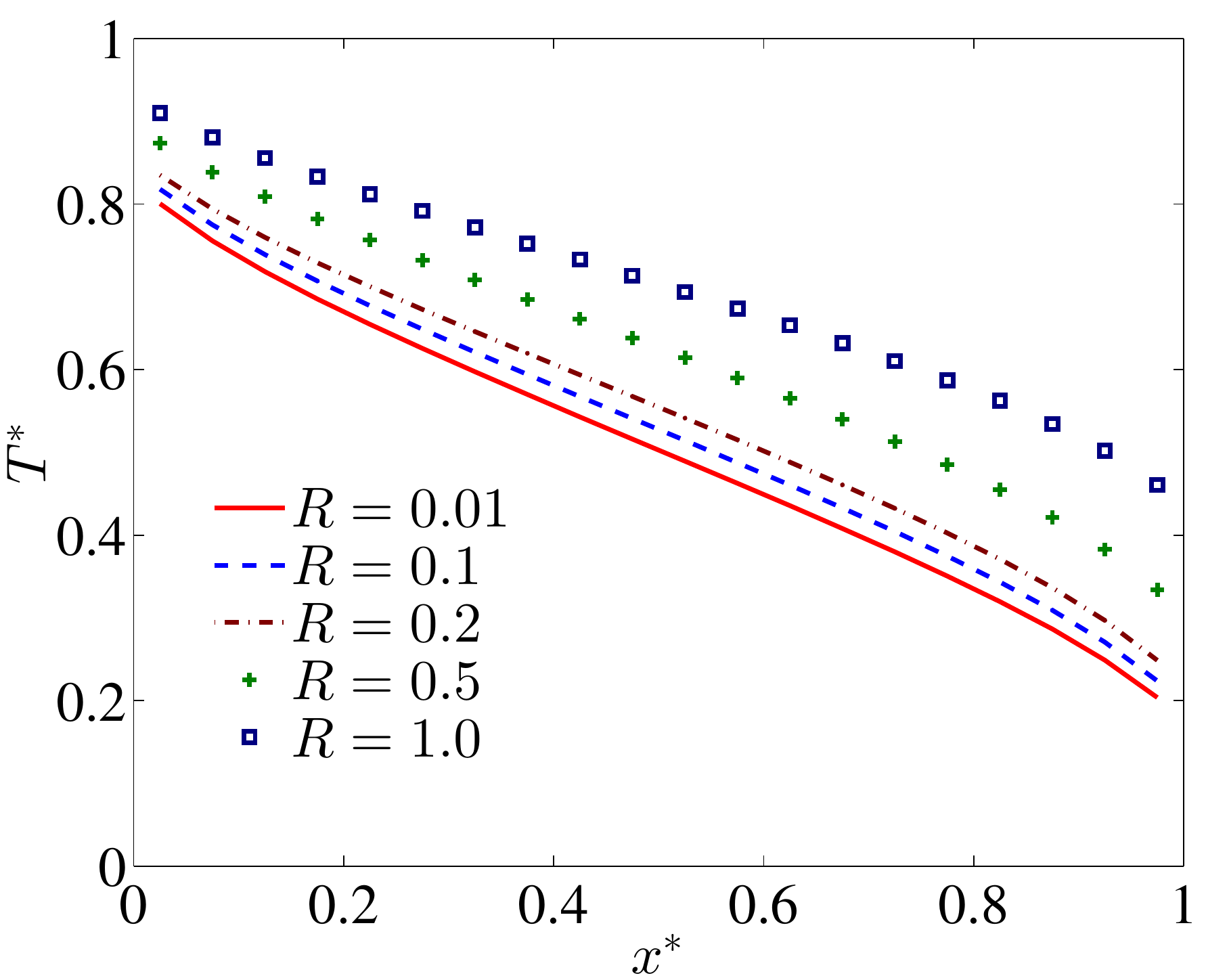}}~~ \\
 \caption{Non-dimensional temperature distribution with different temperature difference, where $x^*=x/L$, $T^*=(T-T_R)/(T_L-T_R)$, $R=\Delta T/T_{0}$, $L=10\mu$m. (a) $T_0=300\text{K}$, (b) $T_0=40\text{K}$. }
 \label{DTfilmL10um}
\end{figure}

In the first part, we fix the film thickness $L=10\mu$m and change $T_0$ from $300\text{K}$ to $40 \text{K}$.
Cases with different temperature difference are simulated and the results are shown in~\cref{DTfilmL10um}, where $0< R \leq 1$.
As $T_0=300\text{K}$, as shown in~\cref{film16}, we have $\overline{\text{Kn}}(T_R)<0.2$, which means the heat transfer is close to that in the diffusive regime.
It can be observed that the temperature jump near the boundaries is not obvious.
As the temperature difference is small enough, i.e., $R=0.01$, the temperature profiles are close to linear.
With the increasing of the temperature difference, the average temperature, which is equivalent to the area among the temperature profile and $x$, $y$ coordinate axis, decreases due to the temperature-dependent thermal conductivity.
The convex temperature profile with $R=1.0$ also indicates that the thermal conductivity decreases with the increasing of the temperature as $250\text{K}<T<500\text{K}$.
As $T_0=40\text{K}$, $\overline{\text{Kn}}(T_0)=7.0$, the thermal transport is in the ballistic regime and the phonon boundary scattering dominates the heat transfer.
As shown in~\cref{film19}, it can be observed that as the temperature difference increases, the temperature jump near the right boundary is strengthen while that on the other boundary is weaken.
That's due to the average Knudsen number is a function of the temperature, i.e., $\overline{\text{Kn}}$ is large near the cold boundary and small near the hot one.

\begin{figure}
 \centering
 \subfloat[]{\label{film11}\includegraphics[width=0.45\textwidth]{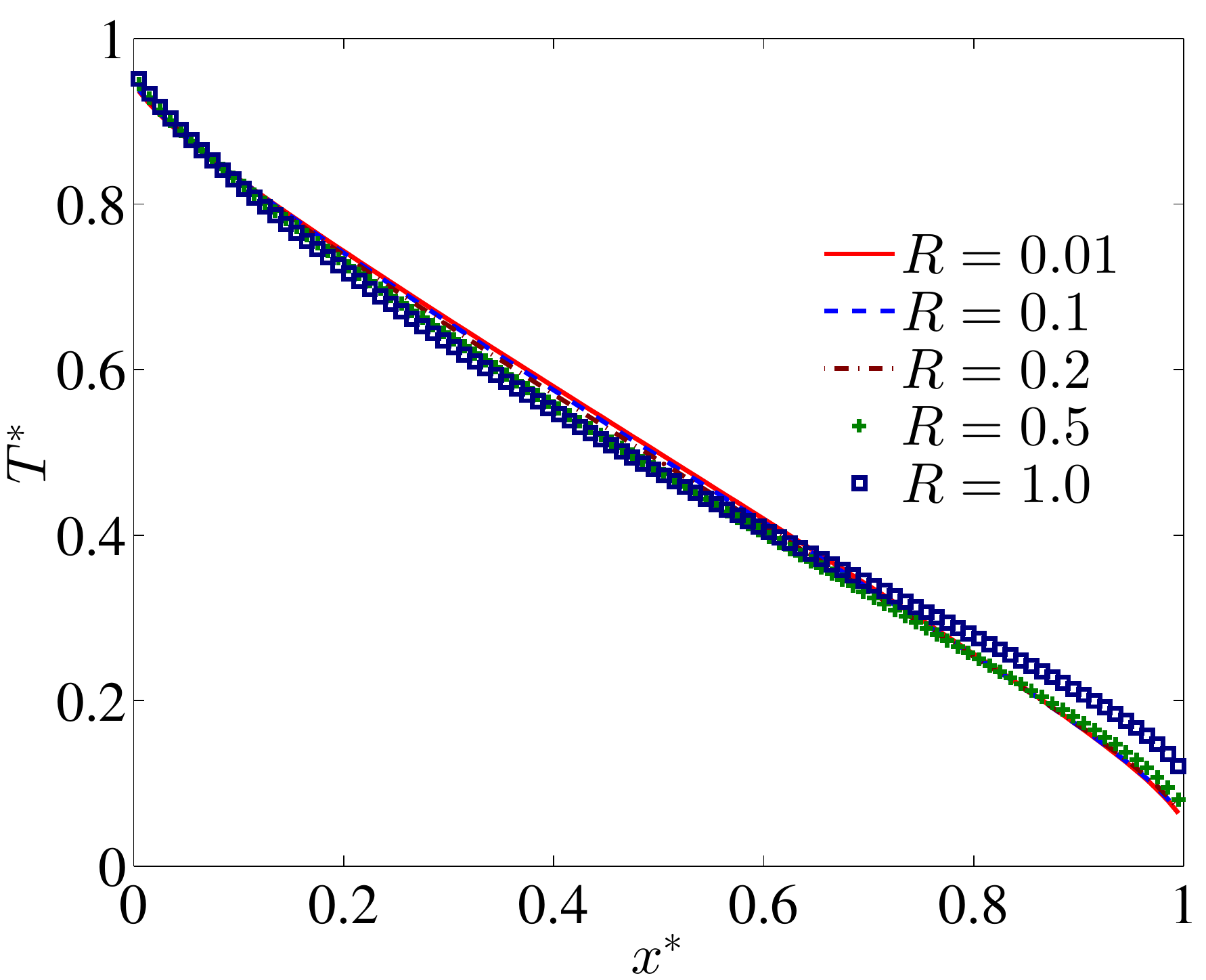}}~~
 \subfloat[]{\label{film17}\includegraphics[width=0.45\textwidth]{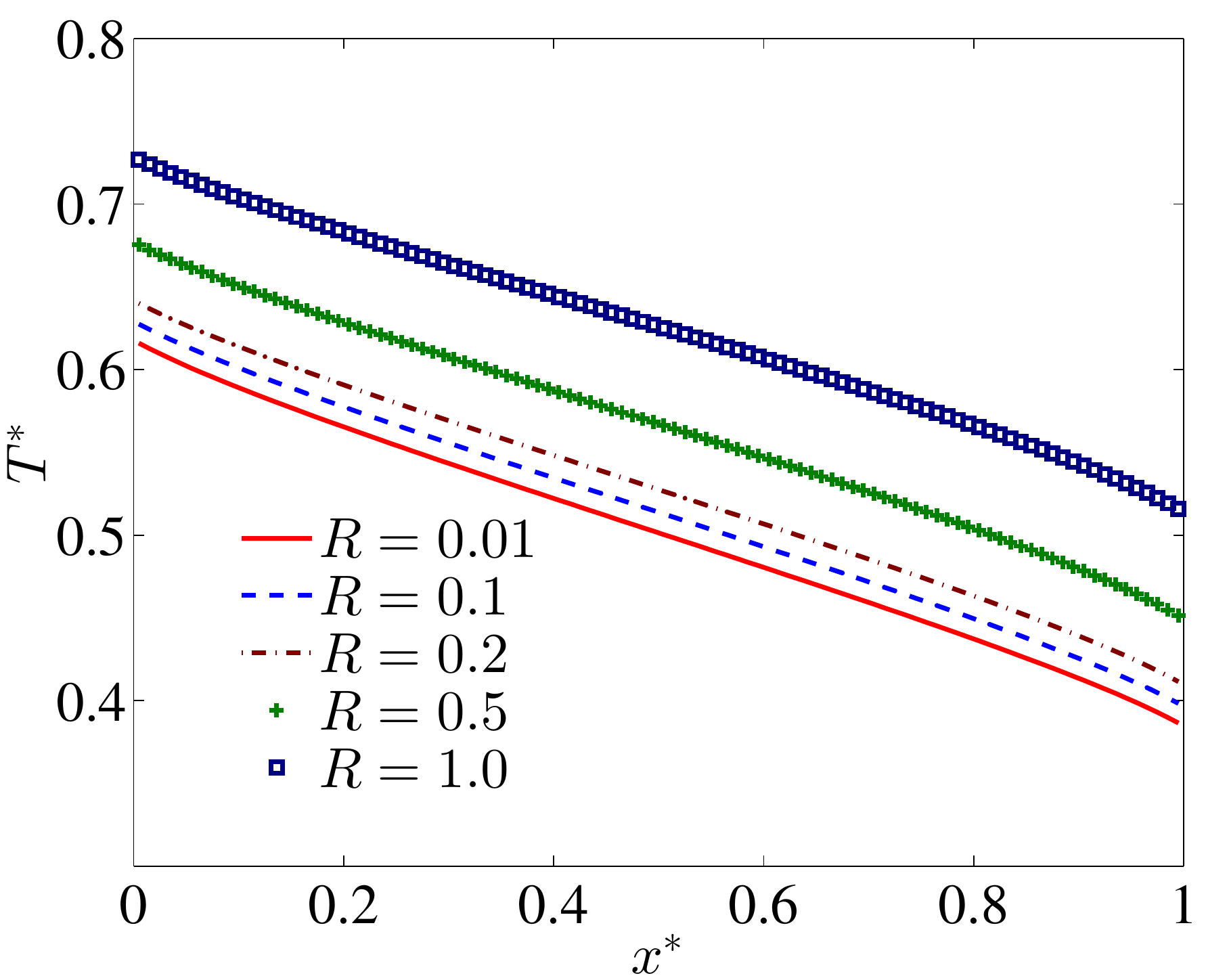}}~~ \\
 \caption{Non-dimensional temperature distribution with different film thickness, where $x^*=x/L$, $T^*=(T-T_R)/(T_L-T_R)$, $R=\Delta T/T_{0}$, $T_0=100\text{K}$. (a) $L=10\mu$m, (b) $L=100$nm. }
 \label{DTfilmT100KL}
\end{figure}

In the second part, we fix $T_0=100\text{K}$ and decrease the film thickness from $L=10\mu$m to $100$nm.
Similar to last part, simulations are made with different temperature difference $0<R \leq 1$.
The temperature distribution at steady state is shown in~\cref{DTfilmT100KL}.
Based on Eq.~\eqref{eq:mfp}, as $T=50\text{K}$, $\bar{\lambda}=31\mu$m, as $T=150\text{K}$, $\bar{\lambda}=1.4\mu$m.
Therefore, as $L=10\mu$m, $0.1 < \overline{\text{Kn}} <4 $, the heat transfer is in the transition regime.
As shown in~\cref{film11}, it can be observed that the deviations of the temperature profiles between $R=0.01$ and $R=1.0$ are small in most areas and the average temperature changes a little as $R$ increases from $0.01$ to $1.0$, which is different from that in the diffusive regime.
In this regime, both the phonon boundary scattering and intrinsic scattering will affect the thermal transport.
The phonon transport in different areas may be in different regimes so that the thermal transport phenomena are complicated.
With the decreasing of the film thickness, $\overline{\text{Kn}}$ increases and the heat transfer comes close to that in the ballistic regime.
As $L=100$nm, from~\cref{film17}, it can be observed that the temperature profiles with different temperature difference are almost parallel to each other.
Because in the interior domain except the area near the boundaries, phonon transport from the hot wall to the cold wall with rare scattering with each other and the thermal conductivity is almost equal everywhere.
As the temperature difference increases, the temperature jump on the boundaries decides the temperature distribution in the interior domain and leads to the increasing of the average temperature, which is similar to that observed in~\cref{film19}.


In summary, as the temperature difference is large, the heat transfer phenomena behave quite differently, which attributes to temperature-dependent phonon mean free path.
As the temperature difference is large, the mean free path may vary significantly even for a given frequency and polarization, so that the heat transfer may be in different regimes.
Therefore, it is important to simultaneously consider the multiscale effects in both the spatial space and frequency space as the temperature difference is large.


\section{Conclusion}

In the present work, a finite-volume discrete unified gas kinetic scheme is developed for multiscale heat transfer with arbitrary temperature difference accounting for the phonon dispersion and polarization.
The temperature at the cell center is updated based on the macroscopic equation according to the macroscopic flux across the cell interface.
The Newtonian method is used to handle the nonlinear relation between the equilibrium state and the temperature at both the cell center and interface.
Several tests of the cross-plane heat transfer show that the present scheme can capture the multiscale phonon transport accurately for large or small temperature difference.
In the diffusive regime, even if the time step or cell size is larger than the relaxation time or mean free path, the transient heat transfer predicted by the present scheme matches well with the solution of the Fourier diffusion equation.
Compared to that under small temperature difference, as the temperature difference is large, the thermal transport phenomena behave quite differently due to the temperature-dependent relaxation time.
With the increasing of the temperature difference, the average temperature in the domain increases in the ballistic regime but decreases in the diffusive regime.
It is important to simultaneously consider the multiscale effects both in the spatial space and frequency space as the temperature difference is large.
The present DUGKS will be a powerful tool to describe the multiscale thermal transport in wide temperature and length ranges.

\section*{APPENDIX: PHONON DISPERSION AND SCATTERING}

The approximate quadratic polynomial dispersions are used to represent the real dispersion relation~\cite{brockhouse1959lattice} of the acoustic phonon branches, i.e.,
\begin{equation}
\omega=c_{1}k+c_{2}k^2,
\label{eq:curves}
\end{equation}
where $c_{1}$, $c_{2}$ are some coefficients.
For LA, $c_{1}=9.01 \times 10^5$cm/s, $c_{2}=-2.0 \times 10^{-3}$$\text{cm}^{2}$/s; for TA, $c_{1}=5.23 \times 10^5$cm/s, $c_{2}=-2.26 \times 10^{-3}$$\text{cm}^{2}$/s~\cite{pop2004analytic}.
For the phonon scattering, the Matthiessen's rule is used to coupled all phonon scattering mechanisms together~\cite{MurthyJY05Review} including the impurity scattering, umklapp (U) and normal (N) phonon-phonon scattering, i.e.,
\begin{equation}
\tau^{-1}=\tau_{{\text{impurity}}}^{-1}+\tau_{{\text{U}}}^{-1}+\tau_{{\text{N}}}^{-1}=\tau_{{\text{impurity}}}^{-1}+\tau_{{\text{NU}}}^{-1},
\label{eq:Matthiessen}
\end{equation}
where the specific formulas of all scattering are shown in Table.~\ref{relaxation}.
\renewcommand\arraystretch{1.5}

\begin{table}
\caption{Phonon scattering~\cite{terris2009modeling}.
 }
\centering
\begin{tabular}{|*{2}{c|}}
 \hline
$\tau_{{\text{impurity}}}^{-1}$   &  $A_{i}\omega^{4}$, ~~$A_{i}=1.498\times10^{-45}~{\text{s}^{\text{3}}}$;       \\
 \hline
LA  & $\tau_{{\text{NU}}}^{-1}=B_{L}\omega^{2}T^{3}$,~~$B_{L}=1.180\times 10^{-24}~{\text{K}^{\text{-3}}}$;      \\
 \hline
\multirow{4}{*}{{\shortstack{TA }}}  & $\tau_{{\text{NU}}}^{-1}=B_T\omega T^4$,~~$0 \leq k <  \pi /a$;      \\
   & $\tau_{{\text{NU}}}^{-1}=B_U\omega^{2}/{\sinh(\hbar\omega/k_{B}T)}$,~~$ \pi /2 \leq k \leq 2\pi /a$;     \\
   & $B_T=8.708\times 10^{-13}~{\text{K}^{\text{-3}}}$,~~ $B_{U}=2.890\times10^{-18}~{\text{s}}$.    \\
   & $a=0.543$nm.    \\
 \hline
\end{tabular}
\label{relaxation}
\end{table}

\section*{References}
\bibliographystyle{IEEEtr}
\bibliography{phonon}

\begin{thebibliography}{10}

\bibitem{cahill2014nanoscale}
D.~G. Cahill, P.~V. Braun, G.~Chen, D.~R. Clarke, S.~Fan, K.~E. Goodson,
  P.~Keblinski, W.~P. King, G.~D. Mahan, A.~Majumdar, {\em et~al.}, ``Nanoscale
  thermal transport. ii. 2003--2012,'' {\em Appl. Phys. Rev.}, vol.~1, no.~1,
  p.~011305, 2014.

\bibitem{cahill2003nanoscale}
D.~G. Cahill, W.~K. Ford, K.~E. Goodson, G.~D. Mahan, A.~Majumdar, H.~J. Maris,
  R.~Merlin, and S.~R. Phillpot, ``Nanoscale thermal transport,'' {\em J. Appl.
  Phys.}, vol.~93, no.~2, pp.~793--818, 2003.

\bibitem{Majumdar98MET}
A.~Majumdar, {\em Microscale energy transport in solids}.
\newblock Taylor and Francis, Washington, DC, 1998.

\bibitem{ju1999microscale}
Y.~S. Ju and K.~E. Goodson, {\em Microscale heat conduction in integrated
  circuits and their constituent films}, vol.~6.
\newblock Springer Science \& Business Media, 1999.

\bibitem{toberer2012advances}
E.~S. Toberer, L.~L. Baranowski, and C.~Dames, ``Advances in thermal
  conductivity,'' {\em Annu. Rev. Mater. Res.}, vol.~42, no.~1, pp.~179--209,
  2012.

\bibitem{Minnich15advances}
A.~J. Minnich, ``Advances in the measurement and computation of thermal phonon
  transport properties,'' {\em J. Phys-condens. Mat.}, vol.~27, no.~5,
  p.~053202, 2015.

\bibitem{zhangZm07HeatTransfer}
Z.~Zhang, {\em Nano/Microscale Heat Transfer}.
\newblock McGraw Hill professional, McGraw-Hill Education, 2007.

\bibitem{ZimanJM60phonons}
J.~M. Ziman, {\em Electrons and phonons: the theory of transport phenomena in
  solids}.
\newblock Oxford University Press, 1960.

\bibitem{ChenG05Oxford}
G.~Chen, {\em Nanoscale energy transport and conversion: a parallel treatment
  of electrons, molecules, phonons, and photons}.
\newblock Oxford University Press, 2005.

\bibitem{srivastava1990physics}
G.~P. Srivastava, {\em The physics of phonons}.
\newblock CRC press, 1990.

\bibitem{hua2015semi}
C.~Hua and A.~J. Minnich, ``Semi-analytical solution to the frequency-dependent
  boltzmann transport equation for cross-plane heat conduction in thin films,''
  {\em J. Appl. Phys.}, vol.~117, no.~17, p.~175306, 2015.

\bibitem{LUO2017970}
X.-P. Luo and H.-L. Yi, ``A discrete unified gas kinetic scheme for phonon
  boltzmann transport equation accounting for phonon dispersion and
  polarization,'' {\em Int. J.Heat Mass Transfer}, vol.~114, no.~Supplement C,
  pp.~970 -- 980, 2017.

\bibitem{MurthyJY12HybridFBTE}
J.~M. Loy, J.~Y. Murthy, and D.~Singh, ``A fast hybrid fourier–boltzmann
  transport equation solver for nongray phonon transport,'' {\em J. Heat
  Transfer}, vol.~135, no.~1, pp.~011008--011008, 2012.

\bibitem{narumanchi2004submicron}
S.~V. Narumanchi, J.~Y. Murthy, and C.~H. Amon, ``Submicron heat transport
  model in silicon accounting for phonon dispersion and polarization,'' {\em
  ASME J. Heat Transfer}, vol.~126, no.~6, pp.~946--955, 2004.

\bibitem{xu_length-dependent_2014}
X.~Xu, L.~F.~C. Pereira, Y.~Wang, J.~Wu, K.~Zhang, X.~Zhao, S.~Bae, C.~T. Bui,
  R.~Xie, J.~T.~L. Thong, B.~H. Hong, K.~P. Loh, D.~Donadio, B.~Li, and
  B.~Özyilmaz, ``Length-dependent thermal conductivity in suspended
  single-layer graphene,'' {\em Nat. Commun.}, vol.~5, p.~3689, 4 2014.

\bibitem{PhysRevB.91.245423}
J.~Cuffe, J.~K. Eliason, A.~A. Maznev, K.~C. Collins, J.~A. Johnson,
  A.~Shchepetov, M.~Prunnila, J.~Ahopelto, C.~M. Sotomayor~Torres, G.~Chen, and
  K.~A. Nelson, ``Reconstructing phonon mean-free-path contributions to thermal
  conductivity using nanoscale membranes,'' {\em Phys. Rev. B}, vol.~91,
  p.~245423, Jun 2015.

\bibitem{PhysRevB.93.235423}
A.~K. Majee and Z.~Aksamija, ``Length divergence of the lattice thermal
  conductivity in suspended graphene nanoribbons,'' {\em Phys. Rev. B},
  vol.~93, p.~235423, Jun 2016.

\bibitem{chavez-angel_reduction_2014}
E.~Chávez-Ángel, J.~S. Reparaz, J.~Gomis-Bresco, M.~R. Wagner, J.~Cuffe,
  B.~Graczykowski, A.~Shchepetov, H.~Jiang, M.~Prunnila, J.~Ahopelto,
  F.~Alzina, and C.~M. Sotomayor~Torres, ``Reduction of the thermal
  conductivity in free-standing silicon nano-membranes investigated by
  non-invasive {Raman} thermometry,'' {\em APL Materials}, vol.~2, p.~012113, 1
  2014.

\bibitem{holland1963analysis}
M.~G. Holland, ``Analysis of lattice thermal conductivity,'' {\em Phys. Rev.},
  vol.~132, pp.~2461--2471, Dec 1963.

\bibitem{armstrong1985n}
B.~H. Armstrong, ``N processes, the relaxation-time approximation, and lattice
  thermal conductivity,'' {\em Phys. Rev. B}, vol.~32, no.~6, p.~3381, 1985.

\bibitem{terris2009modeling}
D.~Terris, K.~Joulain, D.~Lemonnier, and D.~Lacroix, ``Modeling semiconductor
  nanostructures thermal properties: The dispersion role,'' {\em J. Appl.
  Phys.}, vol.~105, no.~7, p.~073516, 2009.

\bibitem{MEHRA201892}
N.~Mehra, L.~Mu, T.~Ji, X.~Yang, J.~Kong, J.~Gu, and J.~Zhu, ``Thermal
  transport in polymeric materials and across composite interfaces,'' {\em
  Appl. Mater. Today}, vol.~12, pp.~92 -- 130, 2018.

\bibitem{laurendeau2005statistical}
N.~Laurendeau, {\em Statistical Thermodynamics: Fundamentals and Applications}.
\newblock Cambridge University Press, 2005.

\bibitem{MazumderS01MC}
S.~Mazumder and A.~Majumdar, ``Monte carlo study of phonon transport in solid
  thin films including dispersion and polarization,'' {\em J. Heat Transfer},
  vol.~123, no.~4, pp.~749--759, 2001.

\bibitem{mittal2010monte}
A.~Mittal and S.~Mazumder, ``Monte carlo study of phonon heat conduction in
  silicon thin films including contributions of optical phonons,'' {\em J. Heat
  Transfer}, vol.~132, no.~5, p.~052402, 2010.

\bibitem{Lacroix05}
D.~Lacroix, K.~Joulain, and D.~Lemonnier, ``Monte carlo transient phonon
  transport in silicon and germanium at nanoscales,'' {\em Phys. Rev. B},
  vol.~72, p.~064305, Aug 2005.

\bibitem{randrianalisoa2008monte}
J.~Randrianalisoa and D.~Baillis, ``Monte carlo simulation of steady-state
  microscale phonon heat transport,'' {\em J. Heat Transfer}, vol.~130, no.~7,
  p.~072404, 2008.

\bibitem{HEATGENERATION11}
B.~T. Wong, M.~Francoeur, and M.~P. Mengüç, ``A monte carlo simulation for
  phonon transport within silicon structures at nanoscales with heat
  generation,'' {\em Int. J.Heat Mass Transfer}, vol.~54, no.~9, pp.~1825 --
  1838, 2011.

\bibitem{ChaiJC93RayEffect}
J.~C. Chai, H.~S. Lee, and S.~V. Patankar, ``Ray effect and false scattering in
  the discrete ordinates method,'' {\em Numer. Heat Transf. Part B}, vol.~24,
  no.~4, pp.~373--389, 1993.

\bibitem{SyedAA14LargeScale}
S.~A. Ali, G.~Kollu, S.~Mazumder, P.~Sadayappan, and A.~Mittal, ``Large-scale
  parallel computation of the phonon boltzmann transport equation,'' {\em Int.
  J. Therm. Sci}, vol.~86, pp.~341 -- 351, 2014.

\bibitem{MurthyJY05Review}
J.~Y. Murthy, S.~V.~J. Narumanchi, J.~A. Pascual-Gutierrez, T.~Wang, C.~Ni, and
  S.~R. Mathur, ``Review of multiscale simulation in submicron heat transfer,''
  {\em Int. J. Multiscale Computat. Eng.}, vol.~3, no.~1, pp.~5--32, 2005.

\bibitem{Pareekshith16BallisticDiffusive}
P.~Allu and S.~Mazumder, ``Hybrid ballistic–diffusive solution to the
  frequency-dependent phonon boltzmann transport equation,'' {\em Int. J.Heat
  Mass Transfer}, vol.~100, no.~Supplement C, pp.~165 -- 177, 2016.

\bibitem{GuoZl13DUGKS}
Z.~Guo, K.~Xu, and R.~Wang, ``Discrete unified gas kinetic scheme for all
  knudsen number flows: Low-speed isothermal case,'' {\em Phys. Rev. E},
  vol.~88, p.~033305, Sep 2013.

\bibitem{GuoZl15DUGKS}
Z.~Guo, R.~Wang, and K.~Xu, ``Discrete unified gas kinetic scheme for all
  knudsen number flows. ii. thermal compressible case,'' {\em Phys. Rev. E},
  vol.~91, p.~033313, Mar 2015.

\bibitem{GuoZl16DUGKS}
Z.~Guo and K.~Xu, ``Discrete unified gas kinetic scheme for multiscale heat
  transfer based on the phonon boltzmann transport equation,'' {\em Int. J.Heat
  Mass Transfer}, vol.~102, pp.~944 -- 958, 2016.

\bibitem{narumanchi2005comparison}
S.~V. Narumanchi, J.~Y. Murthy, and C.~H. Amon, ``Comparison of different
  phonon transport models for predicting heat conduction in
  silicon-on-insulator transistors,'' {\em J. Heat Transfer}, vol.~127, no.~7,
  pp.~713--723, 2005.

\bibitem{RAN2018616}
X.~Ran, Y.~Guo, and M.~Wang, ``Interfacial phonon transport with
  frequency-dependent transmissivity by monte carlo simulation,'' {\em Int.
  J.Heat Mass Transfer}, vol.~123, pp.~616 -- 628, 2018.

\bibitem{brockhouse1959lattice}
B.~N. Brockhouse, ``Lattice vibrations in silicon and germanium,'' {\em Phys.
  Rev. Lett.}, vol.~2, pp.~256--258, Mar 1959.

\bibitem{pop2004analytic}
E.~Pop, R.~W. Dutton, and K.~E. Goodson, ``Analytic band monte carlo model for
  electron transport in si including acoustic and optical phonon dispersion,''
  {\em J. Appl. Phys.}, vol.~96, no.~9, pp.~4998--5005, 2004.

\bibitem{Nanoletter11Hop}
P.~E. Hopkins, C.~M. Reinke, M.~F. Su, R.~H. Olsson, E.~A. Shaner, Z.~C.
  Leseman, J.~R. Serrano, L.~M. Phinney, and I.~El-Kady, ``Reduction in the
  thermal conductivity of single crystalline silicon by phononic crystal
  patterning,'' {\em Nano Lett.}, vol.~11, no.~1, pp.~107--112, 2011.
\newblock PMID: 21105717.

\bibitem{Abramovitch65Math}
M.~Abramowitz and I.~Stegun, {\em Handbook of Mathematical Functions: With
  Formulas, Graphs, and Mathematical Tables}.
\newblock Applied mathematics series, Dover Publications, 1964.

\bibitem{SwebyPK84Fluxlimiter}
P.~K. Sweby, ``High resolution schemes using flux limiters for hyperbolic
  conservation laws,'' {\em Siam J. Numer. Anal.}, vol.~21, no.~5,
  pp.~995--1011, 1984.

\bibitem{LIU2018313}
H.~Liu, Y.~Cao, Q.~Chen, M.~Kong, and L.~Zheng, ``A conserved discrete unified
  gas kinetic scheme for microchannel gas flows in all flow regimes,'' {\em
  Comput. Fluids}, vol.~167, pp.~313 -- 323, 2018.

\bibitem{MajumdarA93Film}
A.~Majumdar, ``Microscale heat conduction in dielectric thin films,'' {\em J.
  Heat Transf}, vol.~115, no.~1, pp.~7--16, 1993.

\bibitem{wangmr17callaway}
Y.~Guo and M.~Wang, ``Heat transport in two-dimensional materials by directly
  solving the phonon boltzmann equation under callaway's dual relaxation
  model,'' {\em Phys. Rev. B}, vol.~96, p.~134312, Oct 2017.

\bibitem{d1986analysis}
E.~R.~G. Eckert and R.~M. Drake~Jr, {\em Analysis Of Heat And Mass Transfer}.
\newblock Taylor \& Francis, 1986.

\bibitem{holman1989heat}
J.~Holman, {\em Heat transfer}.
\newblock Mechanical engineering series, McGraw-Hill, 1989.

\end{thebibliography}

\end{document}